\documentclass[aps,pra,nobalancelastpage,superscriptaddress,twocolumn,reprint]{revtex4-2}
\usepackage{graphicx}
\usepackage{amsmath}
\usepackage{amsthm}
\usepackage{braket}
\usepackage{centernot}
\usepackage[normalem]{ulem}
\usepackage{xcolor}
\definecolor{blueviolet}{rgb}{0.2, 0.2, 0.6}
\definecolor{webgreen}{rgb}{0,.5,0}
\definecolor{webbrown}{rgb}{.6,0,0}
\usepackage[bookmarks=false,
colorlinks=true, 
urlcolor=webbrown, 
linkcolor=blueviolet, 
citecolor=webgreen,
pdfstartpage=1,
pdfstartview={FitH},  
bookmarksopen=false
]{hyperref}
\usepackage{booktabs}
\usepackage[nameinlink,capitalize]{cleveref}

\usepackage{standalone, gensymb, tensor, amsmath,amsfonts,amssymb,amsthm,bbm,bm, braket}
\usepackage[T1]{fontenc}
\usepackage{hyperref}
\usepackage[inline]{enumitem}
\usepackage{scalerel, physics}
\usepackage{wasysym}
\usepackage{comment}
\usepackage{enumerate}
\usepackage{graphicx}
\usepackage{mathtools}
\usepackage{soul}
\usepackage{xparse}
\usepackage{dsfont}
\usepackage{ifthen}
\usepackage{tensor}
\usepackage{tikz}
\usepackage{pgfplots}
\usepackage{tikz-network}
\usepackage{simplewick} 
\usepackage{ mathrsfs }

\usetikzlibrary{patterns,decorations.pathreplacing}
\theoremstyle{break}        

\theoremstyle{break}

\usepackage{color}
\definecolor{alessandrored}{RGB}{250, 153, 140}
\definecolor{OliveGreen}{RGB}{85,107,47}
\definecolor{NavyBlue}{RGB}{0,0,128}
\definecolor{alessandrogreen}{RGB}{17, 127, 35}
\definecolor{jonasgreen}{RGB}{81, 160, 37}
\usepackage{tensor}
\usepackage{xifthen}
\usepackage{xargs}

\definecolor{blue1}{RGB}{0, 0, 255}
\definecolor{blue2}{RGB}{0, 150, 255}
\definecolor{blue3}{RGB}{0, 71, 171}
\definecolor{blue4}{RGB}{100, 149, 237}
\definecolor{blue5}{RGB}{93, 63, 211}

\definecolor{red1}{RGB}{238, 75, 43}
\definecolor{red2}{RGB}{233, 116, 81}
\definecolor{red3}{RGB}{222, 49, 99}
\definecolor{red4}{RGB}{250, 160, 160}
\definecolor{red5}{RGB}{236, 88, 0}

\definecolor{green1}{RGB}{80, 180, 152}
\definecolor{green2}{RGB}{166,218,149}

\definecolor{orange1}{RGB}{255, 116, 23}

\definecolor{grey1}{RGB}{98,98,98}
\definecolor{grey2}{RGB}{211,211,211}
\definecolor{grey3}{RGB}{192,192,192}
\definecolor{grey4}{RGB}{169,169,169}
\definecolor{yellow1}{RGB}{253, 218, 13}
\definecolor{purple1}{RGB}{3, 38, 241}
\definecolor{green3}{RGB}{147, 197, 114}
\definecolor{orange2}{RGB}{255, 170, 51}

\newcommandx{\fineq}[5][1=-.8ex,2=1,3=1,5=0]{
	\begin{tikzpicture}[baseline={([yshift=#1]current  bounding  box.center)}, scale = #2, every node/.style={scale = #3},rotate around={#5:(0,0)},every node/.style={transform shape}]
		#4
	\end{tikzpicture}
}

\definecolor{bertinired}{RGB}{232,102,102}
\definecolor{bertiniblue}{RGB}{101,147,245}
\definecolor{bertinigreyblue}{RGB}{101,147,245}
\definecolor{bertinigreyred}{RGB}{232,102,102}
\definecolor{bertinivioletc}{RGB}{45,130,60}
\definecolor{bertinigreen}{RGB}{166,218,149}
\definecolor{bertiniorange}{RGB}{255, 116, 23}
\definecolor{OliveGreen}{RGB}{85,107,47}
\definecolor{NavyBlue}{RGB}{0,0,128}
\definecolor{bertiniviolet}{RGB}{210,145,178}
\definecolor{bertinigrey1}{RGB}{98,98,98}
\definecolor{bertinigrey2}{RGB}{211,211,211}
\definecolor{bertinigrey3}{RGB}{192,192,192}
\definecolor{bertinigrey4}{RGB}{169,169,169}

\newcommandx{\tikzdiagup}{
	\tikz {\draw[thick] (0,0)--(0.15,0.15); \draw (0,0) rectangle (0.15,0.15);}
}

\newcommandx{\gatecross}[1][1=0.5]{
	\pgfmathparse{#1/2.0}
	\let\x\pgfmathresult
	\draw[thick] (-\x,-\x) -- (\x,\x);
	\draw[thick] (\x,-\x) -- (-\x,\x);
}
\newcommandx{\gatesqu}[2][1=0.25,2=]{
	\pgfmathparse{#1/2.0}
	\let\x\pgfmathresult
	\ifthenelse{\equal{#2}{}}{
		\draw[thick, fill=white, rounded corners=2pt] (-\x,\x) rectangle (\x,-\x);
	}{
		\draw[thick, fill=#2, rounded corners=2pt] (-\x,\x) rectangle (\x,-\x);
	}
}

\newcommandx{\gatemark}[2][1=0.075,2=tr]{
	\pgfmathparse{#1}
	\let\l\pgfmathresult
	\ifthenelse{\equal{#2}{topleft}}{
		\draw[thick] (0,\l) -- ++(-\l,0) --++ (0,-\l);
	}{}
	\ifthenelse{\equal{#2}{topright}}{
		\draw[thick] (0,\l) -- ++(\l,0) --++ (0,-\l);
	}{}
	
	\ifthenelse{\equal{#2}{bottomleft}}{
		\draw[thick] (0,-\l) -- ++(-\l,0) --++ (0,\l);
	}{}
	\ifthenelse{\equal{#2}{bottomright}}{
		\draw[thick] (0,-\l) -- ++(\l,0) --++ (0,\l);
	}{}
	
}

\newcommandx{\squaregate}[3][1=0,2=0,3=white]
{
	\begin{scope}[shift={(#1,#2)},rounded corners= 2pt]
		\draw[thick,fill=#3] (-.13,-.13) rectangle (.15,.15);
	\end{scope}
}

\newcommandx{\roundgate}[6][1=0,2=0,3=1,4=topright,5=white,6=-1]{
	\pgfmathparse{#3}
	\let\l\pgfmathresult
	\begin{scope}[shift={(#1,#2)}]
		\gatecross[\l]
			\pgfmathparse{\l/2.0}
		\let\s\pgfmathresult
		\gatesqu[\s][#5]
		\pgfmathparse{\l*0.15}
		\let\m\pgfmathresult
	\ifthenelse{\equal{#6}{-1}}{		\gatemark[\m][#4]
	}{	\node at ({0},{0}) {\scalebox{1.3}{$#6$}};}
\end{scope}
}

\newcommandx{\wcirc}[2]{\begin{scope}
		\draw[fill=white] (#1,#2) circle (0.15);	\end{scope}} 
\newcommandx{\wcircc}[2]{\begin{scope}
		\draw[fill=white] (#1,#2) circle (0.13);	\end{scope}} 
\newcommandx{\wsqr}[2]{\begin{scope}
		\draw[fill=white,shift={(#1,#2)}] (-.13,.13) rectangle (.13,-.13);	\end{scope}} 
\newcommandx{\wsqrr}[2]{\begin{scope}
		\draw[fill=white,shift={(#1,#2)}] (-.11,.11) rectangle (.11,-.11);	\end{scope}}
\newcommandx{\bcirc}[2]{\begin{scope}
		\draw[fill=black] (#1,#2) circle (0.15);	\end{scope}} 
\newcommandx{\thetastate}[4][1=0,2=0,3=1,4=]{
	\pgfmathparse{#3/2}
	\let\l\pgfmathresult
	\pgfmathparse{\l*0.15}
	\let\m\pgfmathresult
	\begin{scope}[shift={(#1,#2)}]
		\draw[thick] (0,0)--(\l,\l);
		\draw[thick] (0,0)--(-\l,\l);
		\ifthenelse{\equal{#4}{}}{
			\draw[fill=white] (0,0) circle (0.15);
		}{
			\draw[thick, fill=#4] (0,0) circle (0.15);
		}
	\end{scope}
}

\newcommandx{\thetastateflipped}[4][1=0,2=0,3=1,4=]{
	\pgfmathparse{#3/2}
	\let\l\pgfmathresult
	\pgfmathparse{\l*0.15}
	\let\m\pgfmathresult
	\begin{scope}[shift={(#1,#2)}]
		\draw[thick] (0,0)--(\l,-\l);
		\draw[thick] (0,0)--(-\l,-\l);
		\ifthenelse{\equal{#4}{}}{
			\draw[fill=white] (0,0) circle (0.15);
		}{
			\draw[thick, fill=#4] (0,0) circle (0.15);
		}
	\end{scope}
}

\newcommandx{\vertgate}[5][1=0,2=0,3=4,4=bertiniorange,5=topright]
{
	\begin{scope}[shift={(#1,#2)}]
		\ifthenelse{\equal{#3}{1}}{
			\roundgate[0][0][1][#5][#4]
		}{
			\foreach \n[evaluate=\n as \y using {2*\n-2}] in {1,...,#3}{
				\roundgate[0][\y][1][#5][#4]
			}
		}
	\end{scope}
}

\newcommandx{\tsfmatV}[8][1=0,2=0,3=l,4=4,5=tr,6=init,7=bertiniorange,8=topright]{
	\begin{scope}[shift={(#1,#2)}]
		\ifthenelse{\equal{#3}{l}}{
			\pgfmathsetmacro{\flag}{0}
		}{
			\pgfmathsetmacro{\flag}{1}
		}
		
		\foreach \y[evaluate=\y as \x using {mod(\y+\flag,2)}] in {1,...,#4}{
			\roundgate[\x][\y][1][#8][#7]
		}
		\ifthenelse{\equal{#5}{tr}}{
			\foreach \y[evaluate=\y as \x using {mod(\y+\flag,2)}] in {#4}{
				\draw [fill=white] (\x-0.5,\y+0.5) circle (0.15);
				\draw [fill=white] (\x+0.5,\y+0.5) circle (0.15);
			}
		}{}
		\ifthenelse{\equal{#6}{init}}{
			\thetastate[\flag][0][1][#7]
		}{}
	\end{scope}
}

\newcommandx{\leftriangle}[5][1=0,2=0,3=4,4=bertiniorange,5=topright]{
	\begin{scope}[shift={(#1,#2)}]
		\pgfmathsetmacro{\t}{#3}
		\pgfmathsetmacro{\steps}{ceil(\t/2)}
		\foreach \i[evaluate=\i as \x using -\t+2*\i-1, evaluate=\i as \ylim using \t-2*\i+2] in {1,...,\steps}{
			\foreach \y[evaluate=\y as \thisx using {\x+\y-1}] in {1,...,\ylim}{
				\roundgate[\thisx][\y][1][#5][#4][2]
			}
		}
	\end{scope}
}

\newcommandx{\rightriangle}[5][1=0,2=0,3=4,4=bertiniorange,5=topright]{
	\begin{scope}[shift={(#1,#2)}]
		\pgfmathsetmacro{\t}{#3}
		\pgfmathsetmacro{\steps}{ceil(\t/2)}
		\foreach \i[evaluate=\i as \x using -\t+2*\i-1, evaluate=\i as \ylim using \t-2*\i+2] in {1,...,\steps}{
			\foreach \y[evaluate=\y as \thisx using {-\x-\y+1}] in {1,...,\ylim}{
				\roundgate[\thisx][\y][1][#5][#4][-1]
			}
		}
	\end{scope}
}

\newcommandx{\eigenVL}[8][1=0,2=0,3=l,4=5,5=tr,6=init,7=bertiniorange,8=topright]{
	\begin{scope}[shift={(#1,#2)}]
		\pgfmathsetmacro{\t}{#4}
		\leftriangle[0][0][\t][#7][#8]
		
		\ifthenelse{\equal{#6}{init}}{
			\drawinitstate[0][0][l][\t][#7]
		}{}
		
		\ifthenelse{\equal{#5}{tr}}{
			\draw[fill=white] \foreach \x in {0,...,\t} {(\x-0.5-\t,0.5+\x) circle (0.15)};
			\ifthenelse{\equal{#3}{r}}{
				\draw[fill=white] (0.5,\t+0.5) circle (0.15);
			}{}
		}{}
		\ifthenelse{\equal{#5}{parttr}}{
			\draw[fill=white] \foreach \x in {0,...,\t} {(\x-0.5-\t,0.5+\x) circle (0.15)};
		}{}
	\end{scope}
}

\newcommandx{\eigenVR}[8][1=0,2=0,3=l,4=5,5=tr,6=init,7=bertiniorange,8=topright]{
	\begin{scope}[shift={(#1,#2)}]
		\pgfmathsetmacro{\t}{#4}
		\rightriangle[0][0][\t][#7][#8]
		
		\ifthenelse{\equal{#6}{init}}{
			\drawinitstate[0][0][r][\t][#7]
		}{}
		
		\ifthenelse{\equal{#5}{tr}}{
			\draw[fill=white] \foreach \x in {0,...,\t}{(-\x+0.5+\t,0.5+\x) circle (0.15)};
			\ifthenelse{\equal{#3}{l}}{
				\draw[fill=white] (-0.5,\t+0.5) circle (0.15);
			}{}
		}{}
	\end{scope}
}

\newcommandx{\tra}[2][1]{\underset{#1}{\text{tr}}\left[#2\right]}

\newcommandx{\tsfmatDgate}[7][1=0,2=0,3=l,4=4,5=tr,6=bertiniorange,7=topright]
{
	\begin{scope}[shift={(#1,#2)}]
		\ifthenelse{\equal{#3}{l}}{
			\pgfmathsetmacro{\flag}{-1}
		}{
			\pgfmathsetmacro{\flag}{1}
		}
		\pgfmathsetmacro{\t}{#4}
		\foreach \i[evaluate=\i as \x using {\flag*\i}, evaluate=\i as \y using \i] in {1,...,\t}{
			\roundgate[\x][\y][1][#7][#6]
		}
		
		\ifthenelse{\equal{#5}{tr}}{
			\foreach \i[evaluate=\i as \x using {\flag*\i}, evaluate=\i as \y using \i] in {\t}{
				\draw [fill=white] (\x-0.5,\y+0.5) circle (0.15);
				\draw [fill=white] (\x+0.5,\y+0.5) circle (0.15);
			}  
		}{}
	\end{scope}
	
}

\newcommandx{\tsfmatD}[8][1=0,2=0,3=l,4=4,5=tr,6=init,7=bertiniorange,8=topright]{
	\begin{scope}[shift={(#1,#2)}]
		\ifthenelse{\equal{#6}{init}}{
			\thetastate[0][0][1][#7]
		}{}
		
		\ifthenelse{\equal{#3}{l}}{
			\pgfmathsetmacro{\flag}{-1}
		}{
			\pgfmathsetmacro{\flag}{1}
		}
		
		\pgfmathsetmacro{\t}{#4}
		\foreach \i[evaluate=\i as \x using {\flag*\i}, evaluate=\i as \y using \i] in {1,...,\t}{
			\roundgate[\x][\y][1][#8][#7]
		}
		
		\ifthenelse{\equal{#5}{tr}}{
			\foreach \i[evaluate=\i as \x using {\flag*\i}, evaluate=\i as \y using \i] in {\t}{
				\draw [fill=white] (\x-0.5,\y+0.5) circle (0.15);
				\draw [fill=white] (\x+0.5,\y+0.5) circle (0.15);
			}  
		}
		\ifthenelse{\equal{#5}{parttr}}{
			\foreach \i[evaluate=\i as \x using {\flag*\i}, evaluate=\i as \y using \i] in {\t}{
				\draw [fill=white] (\x+0.5*\flag,\y+0.5) circle (0.15);
			}  
		}
		{}
	\end{scope}
}

\newcommandx{\drawinitstate}[5][1=0,2=0,3=l,4=4,5=bertiniorange]{
	\pgfmathsetmacro{\t}{#4}
	\begin{scope}[shift={(#1,#2)}]
		\pgfmathsetmacro{\steps}{ceil((\t-1)/2)}
		\ifthenelse{\equal{#3}{l}}{
			\foreach \i[evaluate=\i as \x using -\t+2*\i] in {0,...,\steps}{
				\thetastate[\x][0][1][#5]
			}
		}{
			\foreach \i[evaluate=\i as \x using -\t+2*\i] in {0,...,\steps}{      
				\thetastate[-\x][0][1][#5]
			}
		}
	\end{scope}
}

\newcommandx{\drawinitstateflipped}[5][1=0,2=0,3=l,4=4,5=bertiniorange]{
	\pgfmathsetmacro{\t}{#4}
	\begin{scope}[shift={(#1,#2)}]
		\pgfmathsetmacro{\steps}{ceil((\t-1)/2)}
		\ifthenelse{\equal{#3}{l}}{
			\foreach \i[evaluate=\i as \x using -\t+2*\i] in {0,...,\steps}{
				\thetastateflipped[\x][0][1][#5]
			}
		}{
			\foreach \i[evaluate=\i as \x using -\t+2*\i] in {0,...,\steps}{      
				\thetastateflipped[-\x][0][1][#5]
			}
		}
	\end{scope}
}

\newcommandx{\eigenDL}[6][1=0,2=0,3=l,4=4,5=bertiniorange,6=topright]{
	\begin{scope}[shift={(#1,#2)}]
		\pgfmathsetmacro{\t}{#4}
		\ifthenelse{\equal{#3}{l}}{
			\eigenVL[0][0][l][\t][tr][init][#5][#6]
			\pgfmathsetmacro{\t}{#4-1}
			\rightriangle[1][0][\t][#5][#6]
			\drawinitstate[1][0][r][\t][#5]
		}{
			\begin{scope}[shift={(-0.5,0.5)}]
				\foreach \i[evaluate=\i as \x using \i, evaluate=\i as \y using \i] in {0,...,\t}{      
					\draw (\x,\y)--++(0.5,0);
					\draw[fill=white] (\x,\y) circle (0.15);
				}
			\end{scope}
		}
	\end{scope}
}

\newcommandx{\eigenDR}[6][1=0,2=0,3=l,4=4,5=bertiniorange,6=topright]{
	\begin{scope}[shift={(#1,#2)}]
		\pgfmathsetmacro{\t}{#4}
		\ifthenelse{\equal{#3}{r}}{
			\eigenVR[0][0][r][\t][tr][init][#5][#6]
			\pgfmathsetmacro{\t}{#4-1}
			\leftriangle[-1][0][\t][#5][#6]
			\drawinitstate[-1][0][l][\t][#5]
		}{
			\begin{scope}[shift={(0.5,0.5)}]
				\foreach \i[evaluate=\i as \x using \i, evaluate=\i as \y using \t-\i] in {0,...,\t}{      
					\draw (\x,\y)--++(0.5,0);
					\draw[fill=white] (\x+0.5,\y) circle (0.15);
				}
			\end{scope}
		}
	\end{scope}
}

\newcommandx{\idonpurity}[2][1=0,2=0]
{
	\begin{scope}[shift={(#1,#2)}]
		\draw[thick] (-0.5,0)--++(-0.1,0.1)--++(0,0.2)--++(0.1,-0.1);
		\draw[thick] (-0.5,0.4)--++(-0.1,0.1)--++(0,0.2)--++(0.1,-0.1);
		\draw[thick] (0.5,0)--++(0.1,0.1)--++(0,0.2)--++(-0.1,-0.1);
		\draw[thick] (0.5,0.4)--++(0.1,0.1)--++(0,0.2)--++(-0.1,-0.1);
	\end{scope}
}

\newcommandx{\swaponpurity}[2][1=0,2=0]
{
	\begin{scope}[shift={(#1,#2)}]
		\draw[thick] (-0.5,0)--++(-0.2,0.2)--++(0,0.6)--++(0.2,-0.2);
		\draw[thick] (-0.5,0.2)--++(-0.075,0.075)--++(0,0.2)--++(0.075,-0.075);
		\draw[thick] (+0.5,0)--++(+0.2,0.2)--++(0,0.6)--++(-0.2,-0.2);
		\draw[thick] (+0.5,0.2)--++(+0.075,0.075)--++(0,0.2)--++(-0.075,-0.075);
	\end{scope}
}

\newcommandx{\hook}[4][1=0,2=0,3=t,4=l]{
	\begin{scope}[shift={(#1,#2)}]
		\ifthenelse{\equal{#3}{t}}{
			\ifthenelse{\equal{#4}{l}}{\draw[thick] (0.5,-0.5) arc (45:-90:0.15);}{\draw[thick] (0.5,-0.5) arc (45:270:0.15);}
		}{\ifthenelse{\equal{#4}{l}}{\draw[ thick] (0.5,-0.5) arc (-45:90:0.15);}{\draw[ thick] (0.5,-0.5) arc (315:90:0.15);}
		}
	\end{scope}
}

\newcommandx{\hhook}[4][1=0,2=0,3=t,4=l]{
	\begin{scope}[shift={(#1,#2)}]
		\ifthenelse{\equal{#3}{t}}{
			\ifthenelse{\equal{#4}{l}}{\draw[thick] (0.5,-0.5) arc (-45:175:0.15);}{\draw[thick] (0.5,-0.5) arc (225:0:0.15);}
		}{\ifthenelse{\equal{#4}{l}}{\draw[ thick] (0.5,-0.5) arc (-45:180:-0.15);}{\draw[ thick] (0.5,-0.5) arc (45:-180:0.15);}
		}
	\end{scope}
}

\newcommandx{\Pproj}[3][3=$P_\Lambda$]{
\begin{scope}[shift={(#1-.5,#2-1)}]
\draw[thick,fill=white] (0,0)rectangle (1,2);
\draw[thick] (0,1.5)--(-.5,1.5);
\draw[thick] (1,1.5)--(1.5,1.5);
\draw[thick] (0,.5)--(-.5,.5);
\draw[thick] (1,.5)--(1.5,.5);
\node[scale=2] at (.5,1) {#3};
\end{scope}}

\definecolor{FcolU}{rgb}{0.71,0.78,0.91}
\definecolor{colLines}{rgb}{0.31,0.31,0.31}
\definecolor{colVMPSLines}{rgb}{0.11,0.11,0.11}
\definecolor{IcolUc}{rgb}{0.71,0.41,0.42}
\definecolor{IcolU}{rgb}{0.71,0.8,0.76}
\definecolor{IcolVMPSc}{rgb}{0.73,0.69,0.7}
\definecolor{IcolVMPS}{rgb}{0.81,0.77,0.78}
\definecolor{colObs}{rgb}{1.,1.,1.}

\def\r{0.08}

\newcommandx{\eightlegs}[2][1=0,2=0]{
	\begin{scope}[shift={(#1,#2)}]
		\foreach \x in {1,...,8}{
			\draw (\x, 0)--++(0,0.25);
			\draw[fill] (\x,0) circle (0.05);
		}
		\foreach \x in {1,3}{
			\pgfmathsetmacro\result{2*\x-1} 
			\node () at (\result,-0.5) {$i_{\x}$};
			\pgfmathsetmacro\result{2*\x}
			\node () at (\result,-0.5) {$j_{\x}$};	
		}
		\foreach \x in {2,4}{
			\pgfmathsetmacro\result{2*\x} 
			\node () at (\result,-0.5) {$i_{\x}$};
			\pgfmathsetmacro\result{2*\x-1}
			\node () at (\result,-0.5) {$j_{\x}$};	
		}
	\end{scope}
}

\newcommandx{\MPSinitialstate}[5][1=0,2=0,3=bertiniorange,4=topright,5=-1]{
\begin{scope}[shift={(#1,#2)},rounded corners=1.5pt]
	\draw[black,thick,fill=#3] 
	(-0.25,.25)--++(.5,0)--++(0,-.3)--++(-.5,0)--cycle;
	\draw[thick] (-.25,.25)--++(-.25,.25);
	\draw[thick] (.25,.25)--++(.25,.25);
	\draw[very thick] (-1,.-.05)--++(2,0);
\ifthenelse{\equal{#5}{-1}}{
	\ifthenelse{\equal{#4}{topright}}{\draw[thick,rounded corners=0.3]
	(-.1,.15)--++(.2,0)--++(0,-0.1); }{}
	\ifthenelse{\equal{#4}{topleft}}{\draw[thick,rounded corners=0.3]
	(.1,.15)--++(-.2,0)--++(0,-0.1); }{}
	\ifthenelse{\equal{#4}{bottomleft}}{\draw[thick,rounded corners=0.3]
	(.1,-.15)--++(-.2,0)--++(0,0.1); }{}	\ifthenelse{\equal{#4}{bottomright}}{\draw[thick,rounded corners=0.3]
	(-.1,-.15)--++(.2,0)--++(0,0.1); }{}}{			\node at ({0},{0.085}) {\scalebox{1.}{{$#5$}}};}
\end{scope}
}

\newcommandx{\Cmatrix}[6][1=0,2=0,3=2,4=bertiniorange,5=,6=topright]{
	\pgfmathsetmacro\result{#3-1} 
	\begin{scope}[shift={(#1,#2)}]
		\foreach \i in {0,...,\result}
		{\foreach \j in {0,...,\i}
			{\roundgate[\i+\j][\i-\j][1][#6][#4]}
		}
		\ifthenelse{\equal{#5}{init}}{
			\foreach \i in {0,...,#3}
			{
				\MPSinitialstate[-1+2*\i][-1][#4]
			}
		}{}
	\end{scope}
}

\renewcommand{\bcirc}{\fineq[-0.5ex][0.7][1]{\cstate[0][0][][black]}}

\newcommandx{\cstate}[4][1=0,2=0,3= ,4=white]{
	\begin{scope}[shift={(0,0)}]
				\draw[fill=#4,thick] (#1,#2) circle (0.13);
				\node[scale=1.1] at (#1,#2) {$#3$};
\end{scope}
}
\newcommandx{\sqrstate}[4][1=0,2=0,3= ,4=white]{
	\begin{scope}[shift={(#1,#2)}]
		\draw[thick,fill=#4] (-0.13,-0.13) rectangle (0.13,0.13) ;
		\node[scale=1.1] at (0,0) {$#3$};
	\end{scope}
}
\newcommandx{\pairproduct}[2][1=0,2=0]
{\begin{scope}[shift={(#1 ,#2)}]
\draw[thick] (-.5,.5) arc(-135:-45:1/1.414);
\sqrstate[0][.5-.1414][][black]	
\end{scope}
}
\newcommandx{\bellpair}[2][1=0,2=0]
{\begin{scope}[shift={(#1 ,#2)}]
		\draw[thick] (-.5,.5) arc(-135:-45:1/1.414);
	\end{scope}
}
\newcommandx{\charge}[3][1=0,2=0,3=black]
{
	\ifthenelse{\equal{#3}{blue}}{\def \chargecolor{bertinigreyblue}
	}{
	\ifthenelse{\equal{#3}{red}}{\def \chargecolor{bertinigreyred}}{\def \chargecolor {#3}}}
\begin{scope}[shift={(#1 ,#2)}]
	\draw[ fill=\chargecolor] circle (0.08);        
\end{scope}
}
\newcommandx{\trianglediag}[7][1=0,2=0,3=1,4=bertiniorange,5= ,6=-1,7=topright]
{\begin{scope}[shift={(#1 ,#2)}]
	\foreach \i in {0,...,#3}
	{	\foreach \j in {0,...,\i}
		{	\roundgate[-\j+2*\i][\j][1][topright][#4][#6]
		}
	}
	\foreach \i in {-1,...,#3}
	{\ifthenelse{\equal{#5}{bellpair}}{\bellpair[\i*2+1][-1]}{\ifthenelse{\equal{#5}{pairproduct}}{\pairproduct[\i*2+1][-1]}{\MPSinitialstate[\i*2+1][-1][#4][#7][#6]}}}
\end{scope}
}
\newcommandx{\projectorleg}[4][1=0,2=0,3=R,4=left]
{
	\begin{scope}[shift={(#1 ,#2)}]
		{\ifthenelse{\equal{#3}{R}}{		\draw[thick] (-.25,-.25)--++(.5,.5);}{\ifthenelse{\equal{#3}{L}}{		\draw[thick] (-.25,.25)--++(.5,-.5);}{\draw[thick] (-.25,0)--++(.5,0);}}
		}
	\draw[thick, fill=white] circle (0.13);
	\ifthenelse{\equal{#4}{right}}{
		\draw[thick] (.0,.07)--++(.07,0)--++(0,-.07);
		\node[scale=0.5] at (0,-.02) {$\alpha$};
	}{}
	\ifthenelse{\equal{#4}{left}}{	
	\draw[thick] (0,.07)--++(-.07,0)--++(0,-.07);
	\node[scale=0.5] at (0,-.03) {$\beta$};
	}{}
	\end{scope}
}
\usetikzlibrary{patterns,decorations.pathreplacing}

\newcommand{\be}{\begin{equation}}
\newcommand{\ee}{\end{equation}}
\newcommand{\ea}{\end{aligned}}
\newcommand{\bea}{\begin{equation}\begin{aligned}}
\newcommand{\eea}{\end{aligned}\end{equation}}

\pgfplotsset{
    colormap={springpastels}{ rgb255=(253, 127, 111) rgb255=(126, 176, 213) rgb255=(178, 224, 97) rgb255=(189, 126, 190) rgb255=(255, 181, 90)  rgb255=(190, 185, 219) rgb255=(253, 204, 229) rgb255=(139, 211, 199)}
    }

\definecolorseries{foo}{hsb}{step}[hsb]{0,1,1}[hsb]{.618,0,0}

\renewcommand{\vec}[1]{\boldsymbol{#1}}

\newtheorem{property}{Property}

\definecolor{bertiniredstronger}{RGB}{200,50,50}

\newcommand{\col}[0]{bertiniviolet}
\newcommand{\rad}[0]{0.25}

\newcommandx{\circletens}[5][1=0,2=0,3=1,4=bertiniviolet,5=0]{
	\begin{scope}[shift={(#1,#2)}]
        \pgfmathparse{#3/2}
        \let\x\pgfmathresult
        \draw[line width=2pt] (-\x, 0) -- (\x, 0);
        \pgfmathparse{#5}
        \ifthenelse{\equal{\pgfmathresult}{1}}
        {
            \draw[thick] (0, 0) -- (0, \x);
        }{
            \ifthenelse{\equal{\pgfmathresult}{2}}
            {
                \draw[thick] (0, 0) -- (0, -\x);
            }{}
        }
        \pgfmathparse{#3 * \rad}
        \let\r\pgfmathresult
        \draw[fill=#4, thick] (0,0) circle (\r);
    \end{scope}
}

\newcommandx{\roundgatewithline}[4][1=0,2=0,3=bertiniviolet,4=0]{
    \begin{scope}[shift={(#1,#2)}]
        \ifthenelse{\equal{#4}{0}}{
            \roundgate[0][0][1][topright][#3]
        }{
        \roundgate[0][0][1][topright][#3][#4]
        }
        \draw[very thick, gray] (-.375,-.375)--(.375,-.375);
        \draw[thick, fill=black] (-.375,-.375) circle (1pt); 
        \draw[thick, fill=black] (.375,-.375) circle (1pt);  
    \end{scope}
}

\newcommandx{\smallcstate}[4][1=0,2=0,3= ,4=white]{
	\begin{scope}[shift={(0,0)}]
				\draw[fill=#4,thick] (#1,#2) circle (0.08);
				\node[scale=1.1] at (#1,#2) {$#3$};
\end{scope}
}

\begin{document}

\title{Solvable Quantum Circuits with non-Markovian Influence Matrices}

\newcommand{\bham}{School of Physics and Astronomy, University of Birmingham, Edgbaston, Birmingham, B15 2TT, UK}

\author{Samuel H. Pickering}
\affiliation{\bham}

\author{Max McGinley}
\affiliation{T.C.M.~Group, Cavendish Laboratory, JJ Thomson Avenue, Cambridge CB3 0HE, UK}

\author{Bhavik Kumar}
\affiliation{Department of Mathematical Sciences, University of Copenhagen, 2100 Copenhagen, Denmark}

\author{Bruno Bertini}
\affiliation{\bham}

\begin{abstract}
Influence matrices encode the action exerted on local subsystems by the rest of an extended quantum many-body system during their evolution. Thus, knowledge of the influence matrix facilitates computationally efficient simulations of local dynamics. Here we propose a new systematic approach to generating quantum circuits with complex dynamics for which the influence matrices can be written down exactly. In contrast to previous frameworks of this kind, such as dual-unitary circuits, the resulting influence matrices are non-Markovian, exhibiting nontrivial temporal correlations. We explicitly construct a broad family of circuits of this kind, based on dressing free-fermion (matchgate) circuits with appropriately chosen interaction terms. We show that, contrary to previous solvable instances, these circuits produce patterns of correlations that closely resemble that of typical many-body systems. Our approach can be directly interpreted in terms of an error correction scheme where the terms breaking the solvability of the influence matrices play the role of errors.    
\end{abstract}

\maketitle

\textit{Introduction.---}The emergence of relaxation in isolated many-body quantum systems can be intuitively understood by recognising that the system acts as a bath for its own finite parts, thereby effectively allowing for information loss~\cite{eisert2015quantum, calabrese2016introduction, serbyn2021quantum, bastianello2022introduction}. This idea can be put on firmer grounds through the concept of the influence matrix~\cite{banuls2009matrix, banuls2009matrix,muellerhermes2012tensor,hastings2015connecting,lerose2020influence}---a self-contained description of the non-unitary evolution of a subsystem induced by the action of the rest of the system. The complexity of simulating local dynamics of quantum many-body systems can then be reduced to the problem of computing the influence matrix itself. Intuitively, influence matrices should be simple when the effective bath produced by the system has weak temporal correlations, i.e., when the dynamics are sufficiently mixing~\cite{lerose2020influence}. 

Recent research has identified a growing number of quantum many-body systems where the influence matrices can be characterised exactly~\cite{bertini2019entanglement, piroli_exact_2020, klobas_2021_exact, klobas2021exact2, giudice2021temporal, prosen2021many, jonay2021triunitary, claeys2024from, yu_hierarchical_2024, bertini2024exact, wang_2025_influence, rampp2025geometric, rampp2025solvable, rampp_2026_hierarchy, pickering2026asymptotically, yang2026solving, klobas2026exact}. These systems, most notably dual-unitary (DU) circuits~\cite{bertini_2019_exact}, have provided the first ever frameworks where one can exactly describe detailed properties of chaotic quantum many-body dynamics and recognise its general structure~\cite{bertini_2019_exact, piroli_exact_2020, bertini2019entanglement, claeys2020maximum, foligno_quantum_2024} (see also Ref.~\cite{bertini2025exactly}). Besides a few isolated examples including certain integrable cellular automata~\cite{klobas_2021_exact, klobas2021exact2, yang2026solving, klobas2026exact} and charge-conserving dual-unitary circuits~\cite{giudice2021temporal, wang_2025_influence}, however, the known systems with exactly solvable influence matrices have Markovian effective baths, i.e.\ the environment acts identically at all times regardless of the previous dynamics, resulting in exceedingly simple spatio-temporal correlations~\cite{bertini2025exactly}. On the contrary, non-Markovian effective baths have been shown to produce different universality classes of complex spatiotemporal correlations~\cite{krajnik2022absence, feldmeier2022emergent, denardis2023nonlinear, gopalakrishnan2024distinct, krajnik2024dynamical}, which can be directly explored in current quantum simulators~\cite{wei2022quantum, rosenberg2024dynamics}. 

While it is in principle possible to describe conditions leading to solvable influence matrices also in the non-Markovian case~\cite{klobas_2021_exact, klobas2021exact2, hubner_kim_2026, insall2026inprep}, these are too complicated to solve directly. As a result, no systematic route for constructing systems with solvable, non-Marovian influence matrices has so far been proposed. In this Letter we provide such a systematic route. We achieve this by starting from free-fermionic quantum circuits and showing that imposing the fermions to be chiral produces Markovian influence matrices. We then demonstrate that the addition of certain fermion scattering terms promotes the influence matrices to non-Markovian whilst retaining solvability. These examples also survive the addition of longitudinal fields, dispersive terms, and terms that completely break the free-fermionic structure. As a result, they show a much richer pattern of correlations than the known Markovian examples. Interestingly, our approach has a direct interpretation in terms of error correction.

\textit{Setup.---}Our focus is on quantum circuits, which model the evolution of many-body systems in discrete spacetime. We consider a 1D array of $q$-level quantum systems (qudits), labelled by $x \in \{0, 1/2, 1, \ldots, L-1/2\}$, each with a local basis $\{\ket{a}_x\}_{a=1}^q$. The circuits we consider have a brickwork structure, generated by a propagator
$\mathbb{U} =\bigotimes_{x=1}^{L} U_{x-1/2,x} \bigotimes_{x=0}^{L-1} U_{x,x+1/2}$ and each $U_{x,y}$ represents a fixed two-qudit unitary gate $U$ applied to qudits $x,y$. For concreteness we choose periodic boundary conditions ($0 \equiv L$), however, we work in the thermodynamic limit $L \rightarrow \infty$ and our results are insensitive to these. 

\begin{figure}
\includegraphics[width=1\columnwidth]{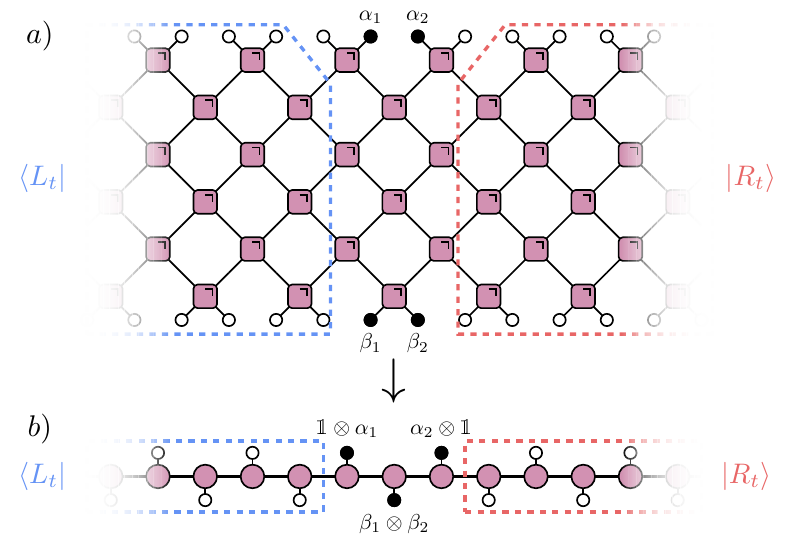}
\caption{Tensor network describing dynamical correlations on top of the infinite temperature state Eq.\eqref{eq:dynamicalcorr} for $x=0$. By representing columns of gates as their own tensors we can compress the notation to convert panel $(a)$ to $(b)$.}
\label{fig:influencematrix}
\end{figure}

Our interest is in local dynamical properties of the Floquet circuit $\mathbb{U}^t$, as exemplified by the infinite-temperature two-point correlation function
\be
\label{eq:dynamicalcorr}
C(x,y,t) =\frac{{\rm tr}[\mathcal O^{(1)}_{x,t} \mathcal O^{(2)}_{0,0}]}{{\rm tr}[I]}, 
\ee
where the operators $\mathcal O^{(i=1,2)}_{x,0}$ act non-trivially on the $d=O(1)$ qudits on the right of $x$, and $\mathcal O^{(i)}_{x,t} \coloneqq \mathbb{U}^{-t} \mathcal O^{(i)}_{x,0} \mathbb{U}^t$ is their Heisenberg-evolved counterpart. Focussing on the case $x=0$, we represent this object graphically as a tensor network in Fig.~\ref{fig:influencematrix}a, with  
\be
\label{eq:gate colors}
\begin{aligned}
U = 
\fineq[-0.8ex][0.8][1]{
   \roundgate[0][0][1][topright][bertinired]
}, \,\,
U^* = 
\fineq[-0.8ex][0.8][1]{
   \roundgate[0][0][1][topright][bertiniblue]
}, \quad
(U\otimes U^*) = 
\fineq[-0.8ex][0.8][1]{
   \roundgate[0][0][1][topright][\col]
}, 
\end{aligned}
\ee

\noindent and where the `circle states' are connections between to forward and backward copies implemented by vectorised operators
\be
\label{eq:bulletstate}
\hspace{-2.5ex}
\vspace{-1ex}
\sum_{a,b=1}^q \!\!\mel*{a}{\mathcal O}{b}\!\frac{\ket*{a}\!\otimes\!\!\ket*{b}}{\sqrt q} = \fineq[0ex][0.6][1]{
    \draw(0, 0)--++(0, .5);
    \cstate[0][0][][]
    \node at (0,-.5) {$\mathcal O$};
    }\,,\,\sum_{a,b=1}^q \!\!\mel*{a}{\mathcal O}{b} \!\frac{\bra*{a}\!\otimes\!\!\bra*{b}}{\sqrt q} = \fineq[-1ex][0.6][1]{
    \draw(0, 0)--++(0, .5);
    \cstate[0][.5][][]
    \node at (0,1) {$\mathcal O$};
    },\!\! 
\ee
with the white circle denoting the special case ${\mathcal O}=I$.

In this setting, the influence matrices $\ket{L_t}$, $\ket{R_t}$ are the states that one would obtain by cutting the circuit tensor network along two verticals, and keeping only the left or right portion, as illustrated in Fig.~\ref{fig:influencematrix}a. These states live on a vertical lattice of $2t+1$ sites, with each site hosting a Hilbert space of dimension $q^2$. They are independent of the observables and thus are an intrinsic property of the circuit. For a given pair of influence matrices $\ket{L_t}$ and $\ket{R_t}$, we can interpret qudits in $[0,d]$ as an open quantum system interacting with a bath (the complementary set of qudits), whose properties are fully captured by $\ket{L_t}$ and $\ket{R_t}$. Therefore, finding an efficient representation of the influence matrices gives direct access to any correlator of the form in Eq.~\eqref{eq:dynamicalcorr}, and indeed higher-order correlations supported in the same spatial region. As well as providing useful physical insight into the thermalization of many-body systems, this procedure can in principle greatly reduce the computational complexity of evaluating dynamical correlations~\cite{banuls2009matrix, banuls2009matrix,muellerhermes2012tensor,hastings2015connecting,lerose2020influence,lerose2023overcoming, lerose2021scaling,magazzu2022feynman,thoenniss2023nonequilibrium,thoenniss2023efficient,ng2023realtime,chen2024grassmann,chen2024realtime,park2024continuous,nayak2025steadystate,sonner2025semigroup, carignano2023temporal,carignano2026itransverse,carignano2025overcoming, yadalam2026process, li2026low}.

\textit{Solvability of influence matrices.---}To describe the influence matrices more succinctly, we will compress our notation by blocking columns of gates into individual tensors of dimension $(q^2 \times q^{2t-1} \times q^{2t-1})^2$, as shown in Fig.~\ref{fig:influencematrix}b. These will be referred to as \textit{column tensors}, and represented by large circles.  Unitarity of the gates $U$ can always be used to reduce the left influence matrix to a contraction of $t-1$ such tensors, namely~\cite{bertini2025exactly} 
\be
\vspace{-1.5ex}
\label{eq:influencematrix}
\begin{aligned}
    \bra*{L_t} = 
\fineq[0.6ex][.8][1]{
    \foreach \i in {0,...,2}{
        \circletens[2*\i][0][1][\col][2]
        \circletens[2*\i+1][0][1][\col][1]
    }
    \foreach \i in {0,...,2}{
        \cstate[2*\i+1][0.6]
        \cstate[2*\i][-0.6]
    }
    \cstate[-.6][0]
    \draw [decorate, thick, decoration = {brace}]   (5.2,-1)--(-.25,-1);
    \node[scale=1.2] at (2.5,-1.5) {$t-1$};
}, 
\end{aligned}
\ee
and similar for $\ket{R_t}$. The horizontal lines represent the collection of $2t-1$ double qudit legs appearing on each side of Fig.~\ref{fig:influencematrix}a, and are accordingly drawn thick. The circle state on the horizontal and vertical bonds are defined as in Eq.~\eqref{eq:bulletstate}, but with $q$ replaced by $q^{2t-1}$ and $q^{2}$ respectively.

Even with this simplification, the dimension of the state in Eq.~\eqref{eq:influencematrix} scales exponentially in time. Therefore, na{\"i}ve approaches to computing $\ket{L_t}$ and $\ket{R_t}$ are prohibitively expensive~\cite{foligno2023temporal}. Nevertheless, in certain cases the influence matrix admits some simplifications. For instance, the key property of dual-unitary circuits can be abstractly stated through the following relations
\be
\label{eq:du_columns}
\begin{aligned}
&\fineq[.4ex][.8][1]{
    \circletens[0][0][1][\col][2]
    \cstate[-.6][0]
    \cstate[0][-.6]
}
=
\fineq[-.8ex][.8][1]{
    \draw[line width=2pt](-.5, 0)--++(.5,0);
    \cstate[-.6][0]
}=
(\fineq[-.8ex][.8][1]{
    \draw[thick](-.5, 0)--++(.5,0);
    \cstate[-.6][0]
})^{\otimes 2t-1}, \\
&\fineq[-1.6ex][.8][1]{
    \circletens[0][0][1][\col][1]
    \cstate[-.6][0]
    \cstate[0][.6]
}
=\fineq[-.8ex][.8][1]{
    \draw[line width=2pt](-.5, 0)--++(.5,0);
    \cstate[-.6][0]
}=
(\fineq[-.8ex][.8][1]{
    \draw[thick](-.5, 0)--++(.5,0);
    \cstate[-.6][0]
})^{\otimes 2t-1},
\end{aligned}
\ee
which imply that the evolution in space is unitary. Applying this to Eq.~\eqref{eq:influencematrix}, the resulting influence matrices are simple product states in time. Thus, $\ket{L_t}$ and $\ket{R_t}$ are efficiently representable, and describe an effective bath with a Markovian property, i.e.~the bath does not mediate temporal correlations.

The simplest generalisation of Eq.~\eqref{eq:du_columns} is
\begin{subequations}
\label{eq:gdu2_columns}
\begin{align}
\label{eq:du2_columns_a}
\fineq[-.8ex][.8][1]{
    \circletens[0][0][1][\col][2]
    \circletens[1][0][1][\col][1]
    \cstate[-.6][0]
    \cstate[0][-.6]
    \cstate[1][.6]
}
&=
\fineq[-1.75ex][.8][1]{
    \circletens[1][0][1][\col][1]
    \cstate[1-.6][0]
    \cstate[1][.6]
}= \fineq[-.8ex][.8][1]{
    \draw[thick](-.5, 0)--++(.5,0);
    \cstate[-.6][0]
}\otimes\left(\fineq[-0.8ex][0.8][1]{
   \roundgate[0][0][1][topright][\col]
   \cstate[-.5][0.5]
   \cstate[-.5][-0.5]
}\right)^{\otimes (t-1)}\hspace{-1cm}, \\
\fineq[-.8ex][.8][1]{
    \circletens[0][0][1][\col][1]
    \circletens[1][0][1][\col][2]
    \cstate[-.6][0]
    \cstate[0][.6]
    \cstate[1][-.6]
}
&=
\fineq[.4ex][.8][1]{
    \circletens[1][0][1][\col][2]
    \cstate[1-.6][0]
    \cstate[1][-.6]
}=\left(\fineq[-0.8ex][0.8][1]{
   \roundgate[0][0][1][topright][\col]
   \cstate[-.5][0.5]
   \cstate[-.5][-0.5]
}\right)^{\otimes (t-1)}\hspace{-1cm}\otimes \fineq[-.8ex][.8][1]{
    \draw[thick](-.5, 0)--++(.5,0);
    \cstate[-.6][0]
}. 
\label{eq:du2_columns_b}
\end{align}
\end{subequations}
These equations are implied by (but are not equivalent to) the so-called DU2 condition~\cite{yu_hierarchical_2024} on the local gates
\begin{align}
    \fineq[-0.8ex][0.8][1]{
   \roundgate[0][0][1][topright][\col]
   \roundgate[1][1][1][topright][\col]
   \cstate[-.5][0.5]
   \cstate[-.5][-0.5]
    \cstate[.5][1.5]
}=\fineq[-0.8ex][0.8][1]{
   \roundgate[1][1][1][topright][\col]
   \draw[thick](.5, -0.5)--++(1,0);
   \cstate[.5][1.5]
   \cstate[.5][0.5]
   \cstate[.5][-0.5]
},\qquad    \fineq[-0.8ex][0.8][1]{
   \roundgate[0][1][1][topright][\col]
   \roundgate[1][0][1][topright][\col]
   \cstate[-.5][0.5]
   \cstate[.5][-0.5]
    \cstate[-.5][1.5]
}=\fineq[-0.8ex][0.8][1]{
   \roundgate[1][0][1][topright][\col]
   \draw[thick](.5, 1.5)--++(1,0);
   \cstate[.5][1.5]
   \cstate[.5][0.5]
   \cstate[.5][-0.5]
}.
\label{eq:local du2}
\end{align}
To distinguish Eqs.~\eqref{eq:gdu2_columns} and Eqs.~\eqref{eq:local du2}, we will refer to the former as column-DU2 (CDU2). 
While the influence matrices of CDU2 circuits are less restricted than DU ones, they still have a product form (with each factor corresponding to a pair of temporal sites), and therefore describe a \textit{Markovian} bath. 

The natural progression of these conditions is 
\begin{subequations}
\label{eq:gdu3_columns}
\begin{align}
\label{eq:gdu3_columns_a}
\fineq[-.8ex][.8][1]{
    \circletens[0][0][1][\col][2]
    \circletens[1][0][1][\col][1]
    \circletens[2][0][1][\col][2]
    \cstate[-.6][0]
    \cstate[0][-.6]
    \cstate[1][.6]
    \cstate[2][-.6]
}
&=
\fineq[-.8ex][.8][1]{
    \circletens[0][0][1][\col][1]
    \circletens[1][0][1][\col][2]
    \cstate[-.6][0]
    \cstate[0][.6]
    \cstate[1][-.6]
}, \\
\fineq[-.8ex][.8][1]{
    \circletens[0][0][1][\col][1]
    \circletens[1][0][1][\col][2]
    \circletens[2][0][1][\col][1]
    \cstate[-.6][0]
    \cstate[0][.6]
    \cstate[1][-.6]
    \cstate[2][.6]
}
&=
\fineq[-.8ex][.8][1]{
    \circletens[0][0][1][\col][2]
    \circletens[1][0][1][\col][1]
    \cstate[-.6][0]
    \cstate[0][-.6]
    \cstate[1][.6]
},
\label{eq:gdu3_columns_b}
\end{align}
\end{subequations}
which, in keeping with the previous nomenclature, we dub column-DU3 (CDU3). These conditions lead to influence matrices composed of two columns of gates, which means that all the temporal sites are non-trivially connected. Therefore, they describe a {\it non-Markovian} bath. At the same time, $\ket{L_t}$ and $\ket{R_t}$ can still be efficiently computed and stored as a matrix product state, with bounded bond dimension $\chi \leq q^2$. Therefore, a solution to these conditions that does not solve any of the previous ones will generically result in influence matrices that are non-Markovian, but still efficiently representable.

A more physical interpretation of the conditions described above can be deduced by considering trajectories of operators in the Heisenberg picture, in the spirit of Feynman paths. We take an orthonormal basis of operators for each site $\{P_{x, \alpha^x}\}_{\alpha^x =0}^{q^2-1}$, such that $P_{x,0} = I_x$ and $\tr\smash{[P_{x,\alpha^x}^\dagger P_{x,\beta^x}]}/\tr\smash{[I_x]} = \delta_{\alpha^x,\beta^x}$ (e.g.~Pauli matrices for ${q = 2}$), and work with a basis of operator strings $\{P_\alpha\}_{\alpha}$, with $\alpha = (\alpha^x)_x$ and $P_\alpha = \bigotimes_x P_{\alpha^x}$.  Then the time-evolved operator $\mathcal{O}_{x,t}$ can be decomposed as
\begin{align}
    \mathcal{O}_{x,t} = \sum_{\alpha_0, \ldots, \alpha_t} c_{\alpha_t}u_{\alpha_t \alpha_{t-1}} \cdots u_{\alpha_{1} \alpha_0} P_{\alpha_0},
    \label{eq:feynman path}
\end{align}
where $u_{\alpha \alpha'} \coloneqq \tr\smash{[P_{\alpha'}^\dagger \mathbb{U}^\dagger P_\alpha \mathbb{U}]}$ and $c_{\alpha} = \tr\smash{[P_\alpha^\dagger \mathcal{O}_{x,0}]}$. Each term in the sum over $\vec{\alpha} = (\alpha_0, \ldots, \alpha_t)$ can be viewed as an operator trajectory that contributes a term $c_{\vec{\alpha}}P_{\alpha_0}$, with amplitude $c_{\vec{\alpha}} \coloneqq c_{\alpha_t}u_{\alpha_t \alpha_{t-1}} \cdots u_{\alpha_{1} \alpha_0}$. The correlator in Eq.~\eqref{eq:dynamicalcorr} can be represented as a sum over the subset of trajectories $\vec{\alpha}$ for which $P_{\alpha_0}$ and $P_{\alpha_t}$ are completely supported within $d$ sites to the right of $0$ and $x$, respectively. This decomposition forms the basis of a number of classical algorithms for approximating operator evolution, in which trajectories involving certain `high-weight' operators are suppressed~\cite{rakovsky2022dissipation,schuster2025, rudolph2026pauli}.

In this language, Eqs.~\eqref{eq:gdu3_columns} are violated if there are operator trajectories with nonzero weight that exhibit a particular kind of `backflow', to use the terminology of Ref.~\cite{rakovsky2022dissipation}. Specifically, if a trajectory (i) begins with support sites $\geq x$, (ii) at some time $t' > 0$ has nontrivial action on site $(x - 3/2)$, and (iii) at time $t > t'$ has support on sites $\geq x$, then this trajectory can contribute to the left hand side of Eq.~\eqref{eq:gdu3_columns_b}, but not to the right hand side. This is because the operator $P_{\alpha_{t'}}$, which is traceless on site $(x - 3/2)$, is annihilated by the contraction with the white circle tensor. Consequently, the suppression of such trajectories is a necessary condition for a circuit to satisfy Eqs.~\eqref{eq:gdu3_columns} (cf.~Fig.~\ref{fig:solvable_trajectories}a).

In Appendix~\ref{sec:operatortrajectories}, we show that a slightly stronger condition on operator backflow, involving higher-order temporal correlations, is also sufficient for Eqs.~\eqref{eq:gdu3_columns} to hold. Since this condition is somewhat more cumbersome to specify, we do not work directly with this formulation. Instead, we use insight from this operator backflow perspective to motivate our search for circuits that constitute nontrivial solutions of Eqs.~\eqref{eq:gdu3_columns}.

\textit{Explicit examples.---}For our search, it is helpful to begin from cases where the Feynman paths can be easily visualised. To this end, we assume that each of our qudits is a collection of $n$ qubits, i.e.\ we set $q=2^n$. We then take our local gates to be free-fermionic, Clifford gates acting on $2n$ qubits. Namely, we construct a set of generators of the Pauli group (identified up to phases), $\{\gamma_{j,x}\}^{x=1/2,\ldots,L}_{j=1,\dots,2n}$, fulfilling the Clifford algebra $\{\gamma_{i,x},\gamma_{j,y}\}=2\delta_{i,j}\delta_{x,y}$, upon which the evolution operator $\mathbb U$ acts as a signed permutation with the signs chosen to conserve fermionic parity $\mathcal P = \prod_{x=1/2}^L \mathcal P_x$ where we set 
\be
\label{eq:Px_definition}
\mathcal P_{x} = \gamma_{1,x} \gamma_{2,x} \cdots \gamma_{2n,x}, \qquad x=1/2, \ldots, L. 
\ee
Using the above definition we can express the Majorana fermions as 
\be
\label{eq:gamma}
\vspace{-1.5ex}
\gamma_{j,x} = {\left[\prod_{y=1/2}^{x-1/2} \mathcal P_y\right]} \ell_{j,x},
\ee
where $\ell_{j,x}$ acts non-trivially, as the local operator $\ell_j$, only at position $x$. Analogously we also define $r_{j,x}= \mathcal P_x \ell_{j,x}$, acting as $r_j$ at site $x$ and as the identity elsewhere. The operators $\ell_j$ and $r_j$ should be understood as the non-trivial heads of the Majorana fermion when its tail runs to the left or right edge of the system respectively.        

If we use Pauli matrices as the operator basis for the decomposition in Eq.~\eqref{eq:feynman path}, then for each choice of initial operator $\mathcal{O}_{x,0} = \gamma_{x,j}$, the evolution follows a single unique trajectory, with $P_{\alpha_\tau}$ equal to a Majorana operator $\gamma_{x(\tau), j(\tau)}$ at each time $\tau \in \{0, 1, \ldots, t\}$. Therefore, we expect the influence matrix to simplify if these trajectories $x(\tau)$ do not turn back on themselves, i.e, $x(\tau)$ is monotonic in $\tau$. Free-fermionic Clifford circuits that obey this condition will be referred to as being \textit{chiral}. This intuition can be formalised in a number of complementary ways: we present the following argument which will generalise most easily beyond free-fermionic Cliffords.

Each gate $U_{x,x+1/2}$  acts on $4n$ Majoranas $\{\gamma_{j,x}\}_{j=1}^{2n}$, $\{\gamma_{j, x+1/2}\}_{j=1}^{2n}$ by permuting them among each other. We categorize these fermions into sets LL, LR, RL, RR, with L and R referring to the left ($x$) and right ($x+1/2$) leg respectively. A fermion is in set XY, with X,Y $\in \{\text{L}, \text{R}\}$ if it is initially located on X and maps to Y. When a fermion of type $j$ hops according to its set XY, its type will generically change to $j'$. Note that, because up to signs $U_{x,x+1/2}$ acts as a permutation on the set of Majoranas, RR and LL (as well as LR and RL) have the same cardinality.
Using now that $U_{x,x+1/2}$ commutes $\mathcal P_{y+1/2}$ as long as $y$ is integer, we can now find pairs of operators such that
\be
\label{eq:freefermcliff_stabilisers}
\begin{aligned}
\sigma_{j}  \fineq[-.8ex][.8][1]{
    \roundgate[0][0][1][topright][bertinired]
    \smallcstate[-.35][.35][][black]
    \node[scale=1.2] at (-1, .35) {$\ell_{j'}$};
    \smallcstate[-.35][-.35][][black]
    \node[scale=1.2] at (-1, -.35) {$\ell_{j}$};
}
&= 
\fineq[-.8ex][.8][1]{
    \roundgate[0][0][1][topright][bertinired]
},
\quad j\in {\rm LL}, \\
\sigma_{k}  \fineq[-.8ex][.8][1]{
    \roundgate[0][0][1][topright][bertinired]
    \smallcstate[.35][.35][][black]
    \node[scale=1.2] at (1, .35) {$r_{k'}$};
    \smallcstate[.35][-.35][][black]
    \node[scale=1.2] at (1, -.35) {$r_{k}$};
}
&= 
\fineq[-.8ex][.8][1]{
    \roundgate[0][0][1][topright][bertinired]
},
\quad k\in {\rm RR}. 
\end{aligned}
\ee
for $\sigma_j=\pm$. The operators $\sigma_{j} (\ell_{j'}\otimes \ell_{j})$  and $\sigma_{k} (r_{k'}\otimes r_{k})$ can be interpreted as (left and right) stabilizers of the gate acting `in space', i.e., from left to right.

These gate stabilizers can be used to characterize the influence matrix in an informative way. In Appendix~\ref{sec:property1}, we use Eqs.~\eqref{eq:freefermcliff_stabilisers} to show
\be
\label{eq:freefermcliff_column_expansion}
\begin{aligned}
\fineq[-1.8ex][.8][1]{
    \circletens[0][0][1][\col][1]
    \cstate[-.6][0]
    \cstate[0][.6]
}=
\fineq[.8ex][.8][1]{
    \circletens[0][0][1][\col][2]
    \cstate[-.6][0]
    \cstate[0][-.6]
}  
=
\fineq[-.6ex][.8][1]{
    \draw[very thick] (-.5,0) -- (.5,0);
    \cstate[-.5][0]
}  
+
\sum_{k} \sigma_k
\fineq[-.6ex][.8][1]{
    \draw[very thick] (-.5,0) -- (.5,0);
    \cstate[-.5][0][][black]
	\node[scale=1.1] at (-1,-.05)  {$R_k$};
}, 
\end{aligned}
\ee
where $\sigma_k=\pm$ and $R_k$ are vectorised non-identity operators (cf.\ Eq.~\eqref{eq:bulletstate}) that belong to a group locally generated by the pairs $(r_{k'_i}\otimes r_{k_i})$, with ${k_i}\in {\rm RR}$, positioned at each gate on the time lattice.

As a result, the CDU2 condition [Eqs.~\eqref{eq:gdu2_columns}] is fulfilled if the contraction of each $R_k$ with the left leg of the next column vanishes.
The left leg of this second column is itself stabilized by a set of operators $L_j$, which are generated by LL-type fermions according to Eq.~\eqref{eq:freefermcliff_stabilisers}. In particular, this implies
\be
\label{eq:unwanted}
\begin{aligned}
\fineq[-1.6ex][.8][1]{
    \circletens[0][0][1][\col][1]
    \cstate[-.6][0][][]
    \cstate[0][.6]
    \node[scale=1.1] at (-1,-.05)  {$R_k$};
}  
=
\fineq[-1.6ex][.8][1]{
    \circletens[0][0][1][\col][1]
    \cstate[-.6][0][][]
    \cstate[0][.6]
    \node[scale=1.1] at (-1.5,-.05)  {$L_j R_k L_j$};
},\\ 
\fineq[.8ex][.8][1]{
    \circletens[0][0][1][\col][2]
    \cstate[-.6][0][][]
    \cstate[0][-.6]
    \node[scale=1.1] at (-1,-.05)  {$R_k$};
}  
=
\fineq[.8ex][.8][1]{
    \circletens[0][0][1][\col][2]
    \cstate[-.6][0][][]
    \cstate[0][-.6]
    \node[scale=1.1] at (-1.5,-.05)  {$L_j R_k L_j$};
},
\end{aligned}
\ee
where $L_j$ is any product of the LL stabilizers in Eq.~\eqref{eq:freefermcliff_stabilisers} and thus a Pauli string. Therefore, if there exists such an $L_j$ that anticommutes with $R_k$, the l.h.s.\ Eq.~\eqref{eq:unwanted} vanishes and Eq.~\eqref{eq:gdu2_columns} is satisfied. The anti-commutation between $L_j$ and $R_k$ is connected with the chirality of the fermions by the following property (proven in Appendix~\ref{sec:property1}) 

\begin{property}
\label{prop:chiralfreefermCliff}
In chiral free-fermionic Clifford circuits, for each $R_k$ appearing in Eq.~\eqref{eq:freefermcliff_column_expansion} there exists an anticummuting left-stabilizer $L_j$. Therefore, these circuits satisfy CDU2 [Eqs.~\eqref{eq:gdu2_columns}]. 
\end{property}

Interestingly, although the above property ensures that chiral, free fermionic Clifford circuits fulfil Eq.~\eqref{eq:gdu2_columns}, they generically do not fulfil the DU conditions in Eqs.~\eqref{eq:du_columns}, nor the local DU2 condition in Eq.~\eqref{eq:local du2}. This is due to the fact that, in general, a chiral trajectory of Majorana operators can remain at the border between system and environment for arbitrarily many time steps, before eventually moving into the environment: this violates Eq.~\eqref{eq:local du2}.

\textit{Beyond free-fermionic Clifford.---}Since the CDU2 conditions arise as a consequence of the mutual algebra between the generators of the left and right stabiliser groups, we can deform these circuits in a way that breaks both the free-fermionic and the Clifford property, while leaving  these generators invariant. The resulting circuit will then continue to fulfil Eq.~\eqref{eq:gdu2_columns}. This is expressed by the following property (proven in Appendix~\ref{sec:property2}) 
\begin{property}
\label{prop:cdu2_general}
Consider a circuit generated by gates of the form $W_{x,x+1}= (u_1\otimes u_2)\,U_{x,x+1} \,(u_3\otimes u_4)$, where $U_{x,x+1}$ is a chiral, free fermionic Clifford gate and $u_{1}$, $u_{2}$, $u_{3}$ and $u_{4}$ are single-qudit unitaries contained in the algebras generated by $\{r_{k'}\}_{k \in \text{RL}}$, $\{\ell_{j'}\}_{j \in \text{LR}}$, $\{r_{j}\}_{j\in {\rm LR}}$ and $\{\ell_{k}\}_{k \in \text{RL}}$ respectively. Then, Eqs.~(\ref{eq:freefermcliff_column_expansion}, \ref{eq:unwanted}) hold, and for each $R_k$ there exists an anticommuting $L_j$. Therefore, this circuit satisfies CDU2 [Eqs.~\eqref{eq:gdu2_columns}].
\end{property}

\begin{figure*}[t]
    \centering
     \includegraphics[width=1\textwidth]{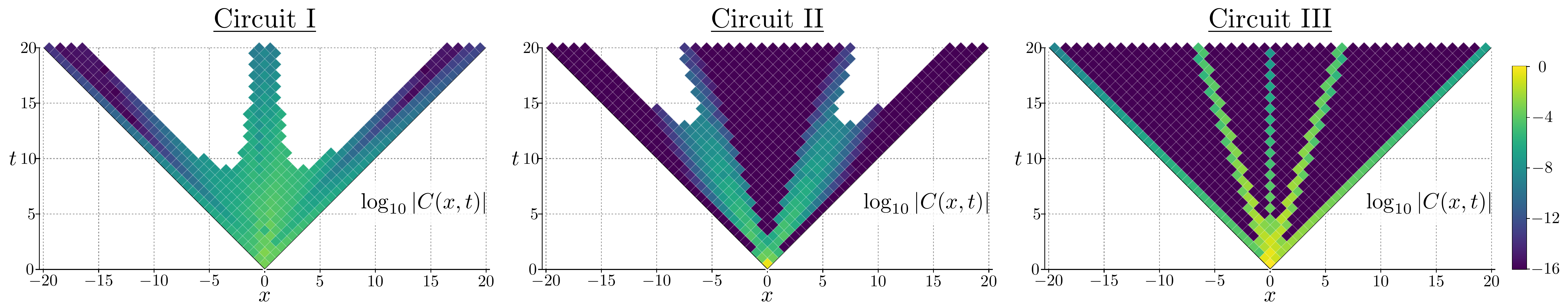}
    \caption{Dynamical correlation functions $C(x,t)$ (between random Hermitian operators acting on two sites) for three example circuits satisfying CDU2/CDU3, displaying varying levels of solvability. The details of each circuit are described in Table \ref{tab:example-circuits}. Numerics are performed using MPS-based simulation of the tilted influence matrix with no truncation error (see End Matter); blank entries indicate that the bond dimension surpassed the cutoff $\chi_c = 384$. The circuits I-III are described in Appendix \ref{sec:tableau} with the general summary being that I and III are CDU2 whilst II is CDU3. The ordering of the circuits corresponds to the degree of solvability beyond that guaranteed by the CDU$n$ conditions (see End Matter).
    } 
    \label{fig:correlations}
\end{figure*}

The unitary perturbations $u_{i}$ ($i = 1,\ldots, 4$) can be written as $e^{i\theta h_i}$ where $h_i$ is a Hermitian element of the appropriate algebra; this is evidently a continuous family, which breaks the Clifford property. Allowed terms in $h_i$ include Majorana bilinears (producing chirality-preserving dispersion between the fermions) as well as local Paulis that are non-local with respect to the Majoranas (which break the free-fermion solvability of the circuit, akin to the longitudinal field in the Ising model), along with further higher-order terms.

\textit{Generalizing to CDU3.---}Let us now move away from CDU2 and seek truly non-Markovian influence matrices fulfilling CDU3. In this case our strategy is to still impose Eq.~\eqref{eq:freefermcliff_column_expansion} but replace Eqs.~\eqref{eq:unwanted} with 
\be
\label{eq:unwanted2}
\begin{aligned}
\fineq[-.8ex][.8][1]{
    \circletens[-1][0][1][\col][1]
    \circletens[0][0][1][\col][2]
    \cstate[-1.6][0][][]
    \cstate[-1][.6]
    \cstate[0][-.6]
    \node[scale=1.1] at (-2,-.05)  {$R_k$};
}  
=
\fineq[-.8ex][.8][1]{
    \circletens[-1][0][1][\col][1]
    \circletens[0][0][1][\col][2]
    \cstate[-1.6][0][][]
    \cstate[-1][.6]
    \cstate[0][-.6]
    \node[scale=1.1] at (-2.5,-.05)  {$L_j R_k L_j$};
},\\
\fineq[-.8ex][.8][1]{
    \circletens[-1][0][1][\col][2]
    \circletens[0][0][1][\col][1]
    \cstate[-1.6][0][][]
    \cstate[0][.6]
    \cstate[-1][-.6]
    \node[scale=1.1] at (-2,-.05)  {$R_k$};
}  
=
\fineq[-.8ex][.8][1]{
    \circletens[-1][0][1][\col][2]
    \circletens[0][0][1][\col][1]
    \cstate[-1.6][0][][]
    \cstate[0][.6]
    \cstate[-1][-.6]
    \node[scale=1.1] at (-2.5,-.05)  {$L_j R_k L_j$};
},
\end{aligned}
\ee
where now $L_j$ are stabilizers of the two-column tensor. In this way, we obtain a solution of CDU3 whenever we can find, for each $R_k$, an $L_j$ that anticommutes with it. Since we are seeking solutions that do not solve CDU2, we must ensure that the two-column tensor possesses stabilizers $L_j$ that are not present for the one-column case, so that Eqs.~\eqref{eq:unwanted} do not hold.

We identify a broad family of such circuits, a representative example of which is constructed as follows. We choose a chiral, free fermionic Clifford circuit acting on ${n=4}$ qubits per site, such that the trajectories of Majorana fermions moving at average velocities $v\in\{-1, -1/3, 1/3, 1\}$. By this we mean that the four Majoranas per site can be labelled as $\{\gamma_{1,x}^{(1)},\gamma_{1,x}^{(1/3)},\gamma_{2,x}^{(1/3)},\gamma_{3,x}^{(1/3)}\}$, such that the circuit satisfies
\begin{align}
    &\mathbb{U}^{-3} \gamma^{(1)}_{1,x} \mathbb{U}^{3} = \gamma^{(1)}_{1,x \pm 3} &&  &\mathbb{U}^{-3} \gamma^{(1/3)}_{i,x} \mathbb{U}^{3} = \gamma^{(1/3)}_{i,x \pm 1},
\end{align}
where $+$($-$) corresponds to $2x+i$ even (odd). This system fulfils Eq.~\eqref{eq:freefermcliff_column_expansion} and Property~\ref{prop:chiralfreefermCliff} ensures that for each of the $R_k$s we can find an anti-commuting $L_j$, i.e., the CDU2 condition is satisfied.  

We now deform the local gate as $U\mapsto UI(\theta)$, where $I(\theta)$ describes an interaction of the form 
\be
\label{eq:pert}
\!\!I(\theta)_{x,x+1/2} = \exp(i\,\theta\,\gamma_{1,x}^{(1)}\gamma_{1,x}^{(1/3)}\gamma_{1,x+1/2}^{(1/3)}\gamma_{1,x+1/2}^{(1)}),
\ee
which generates $1\leftrightarrow3$ scattering processes as  
\begin{align}
\hspace{-1ex}
\fineq[0ex][.7][1]{
    \draw[thick, ->] (-2,-1.4)--++(1.4,0);
    \draw[thick, ->] (-2,-1.42)--++(0,1.4);
    \node[scale=1.2] at (-1,-1.8) {$x$};
    \node[scale=1.2] at (-2.3,-.8) {$t$};
    \draw[very thick] (-.9,-.9)--(0,0);
    \draw[very thick] (0,0)--(-1,1);
    \draw[very thick] (0,0)--(-1/3,1);
    \draw[very thick] (0,0)--(1/3,1);
    \smallcstate[0][0][][]
    \node[scale=1.2] at (-1.1,-1.1) {$1$};
    \node[scale=1.2] at (-1.2,1.2) {$-1$};
    \node[scale=1.2] at (-1.2/3,1.3) {$-\frac{1}{3}$};
    \node[scale=1.2] at (1.2/3,1.3) {$\frac{1}{3}$};
} + \hspace{-2ex}
\fineq[-.8ex][.7][1]{
    \draw[very thick] (-.9/3,-.9)--(0,0);
    \draw[very thick] (0,0)--(-1,1);
    \draw[very thick] (0,0)--(-1/3,1);
    \draw[very thick] (0,0)--(.9,.9);
    \smallcstate[0][0][][]
    \node[scale=1.2] at (-1.15/3,-1.15) {$\frac{1}{3}$};
    \node[scale=1.2] at (-1.2,1.2) {$-1$};
    \node[scale=1.2] at (1,1) {$1$};
    \node[scale=1.2] at (-1.2/3,1.3) {$-\frac{1}{3}$};
} + \hspace{-2ex}
\fineq[-.8ex][.7][1]{
    \draw[very thick] (.9/3,-.9)--(0,0);
    \draw[very thick] (0,0)--(-1,1);
    \draw[very thick] (0,0)--(1/3,1);
    \draw[very thick] (0,0)--(1,1);
    \smallcstate[0][0][][]
    \node[scale=1.2] at (1.1/3,-1.1) {$-\frac{1}{3}$};
    \node[scale=1.2] at (-1.2,1.2) {$-1$};
    \node[scale=1.2] at (1.2/3,1.3) {$\frac{1}{3}$};
} + \hspace{-1ex}
\fineq[-.8ex][.7][1]{
    \draw[very thick] (.9,-.9)--(0,0);
    \draw[very thick] (0,0)--(1,1);
    \draw[very thick] (0,0)--(-1/3,1);
    \draw[very thick] (0,0)--(1/3,1);
    \smallcstate[0][0][][]
    \node[scale=1.2] at (1.1,-1.1) {$-1$};
    \node[scale=1.2] at (1.2,1.2) {$1$};
    \node[scale=1.2] at (-1.2/3,1.3) {$-\frac{1}{3}$};
    \node[scale=1.2] at (1.2/3,1.3) {$\frac{1}{3}$};
}
\notag
\end{align}
among the Majoranas. The deformation $I(\theta)$ goes beyond the local rotations $u_1 \otimes u_2$ described in Property~\ref{prop:cdu2_general}.
In this case (see Appendix~\ref{sec:property3}) the expansion in Eq.~\eqref{eq:freefermcliff_column_expansion} essentially still holds (with $\sigma_j$ replaced by appropriate $\theta$-dependent coefficients). However, there are fewer left-stabilizers for each gate, and thus we lose Property \ref{prop:cdu2_general} for the one-column tensor. Crucially, the form of the perturbation ensures that these lost left-stabilizers are recovered when two columns are concatenated together, which gives us the following.

\begin{property}
\label{prop:cdu3_general}
Consider the local gate $U I(\theta)$, where $U$ is the chiral Majorana gate described above with velocities $v \in \{-1,-1/3,1/3,1\}$, and $I(\theta)$ as in Eq.~\eqref{eq:pert}.
Then for $\theta \notin (\pi/2)\mathbb{Z}$, the corresponding circuit fulfils the CDU3 condition \eqref{eq:gdu3_columns} but not CDU2  \eqref{eq:gdu2_columns}. The same holds for the gate $(u_1\otimes u_2)\, U I(\theta)$, where $u_1,u_2$ are as defined in Property~\ref{prop:cdu2_general}, but satisfy additional constraints (see App.~\ref{sec:property3}).
\end{property}
In Appendix \ref{sec:property3}, we describe a much more general family of chiral circuits and corresponding perturbations that obey the same property---these have velocities $v \in [-1,-1/T] \cup \{0\} \cup [1/T,1]$ for some odd integer $T$.
 The form of these constructions is reminiscent of DU gates on qubits~\cite{bertini_2019_exact}, where the core structure is given by the product of a Clifford gate supporting fermions with velocities $v\in\{-1,1\}$ and the exponential of a quartic term in the Majoranas. 
Therefore, the models that we have introduced can be thought of as a generalisation of these DU circuits, where we allowed the unperturbed Clifford free fermions to have arbitrary velocities.

\textit{Interpretation via quantum codes.---}Eqs.~(\ref{eq:unwanted}, \ref{eq:unwanted2}) point to a natural interpretation of Properties \ref{prop:chiralfreefermCliff}--\ref{prop:cdu3_general} in terms of quantum error-correcting codes. 
We can view the circuit in the crossed channel, i.e.\ evolving right-to-left, by interpreting each (single-copy) column tensor as an operator $V_{TL \leftarrow R}$, mapping states $\ket{\psi}_R$ on the right leg ($R$) to states on the left ($L$) and vertical ($T$) legs.  $V_{TL \leftarrow R}$ will be interpreted as an encoding map, relating `logical' information in $R$ to `physical' states on $TL$. Dynamics is then generated by concatenating these maps, column-by-column, such that the output $L$ of of the $n$th step becomes the input $R$ of the $(n+1)$th step [cf.~Eq.~\eqref{eq:influencematrix}].

The DU condition in Eqs.~\eqref{eq:du_columns} is equivalent to $V_{TL \leftarrow R}$ being isometric, and hence information-preserving. Conversely, for non-DU circuits, $V_{TL \leftarrow R}$ corrupts the information on $R$ nontrivially. Nevertheless, the existence of left (LL) stabilizers for each gate [Eq.~\eqref{eq:freefermcliff_stabilisers}] implies that the image of $V_{TL \leftarrow R}$ lies in a subspace $W_L \subset \mathcal{H}_L \cong (\mathbb{C}^q)^{\otimes({2t+1})}$. When we apply the next step of the sideways-evolving dynamics [as described by the operators $R_j$, Eq.~\eqref{eq:freefermcliff_column_expansion}], this could in principle corrupt the logical information on $R$ further. However, Properties \ref{prop:chiralfreefermCliff} and \ref{prop:cdu2_general} ensure that each $R_j$ is a \textit{correctable error} for the code space~\cite{knill2002introduction}---that is, they satisfy the Knill-Laflamme conditions $\Pi_L R_k^\dagger R_{k'} \Pi_L \propto \Pi_L$ ($\Pi_L$ projector onto $W_L$) \footnote{The anticommutation of $L_j$ and $R_k$ in Eq.~\eqref{eq:unwanted} makes it clear that the conditions for $R_k$ to be a \textit{detectable} error are satisfied $\Pi_L R_{k} \Pi_L \propto \Pi_L$ (Ref.~\cite{knill2002introduction}). Correctability then follows from the closure of $\{R_k\}_k$ under products.}. 
Thus, even though we are applying a non-isometric map $V_{TL \leftarrow R}$ in the second step, it is information-preserving when applied to the output of the first step, which lies in the code space $W_L$. This emergent unitarity leads to the simplification in Eq.~\eqref{eq:gdu2_columns}.

For CDU3 circuits that are not CDU2, the second step of the sideways evolution generated by $V_{TL \leftarrow R}$ fails to be information-preserving. Instead, after two steps the state will lie in a smaller subspace $W_L^{(2)} \subset W_L$. Because $W_L^{(2)}$ has a greater number of stabilizers $L_j$, it can protect against a wider range of errors, and thus lead to emergent unitarity. 
More generally, a circuit whose sideways evolution is information-preserving for steps $n, n+1, \ldots$ will obey a CDU$n$ property, where the left and right hand sides of Eq.~\eqref{eq:gdu3_columns} have $n$ and $(n-1)$ tensors, respectively. The integer $n$ plays a role akin to the  initialization time of a Floquet code~\cite{hastings2021dynamically}.

\begin{figure*}[t]
    \centering
     \includegraphics[width=1\textwidth]{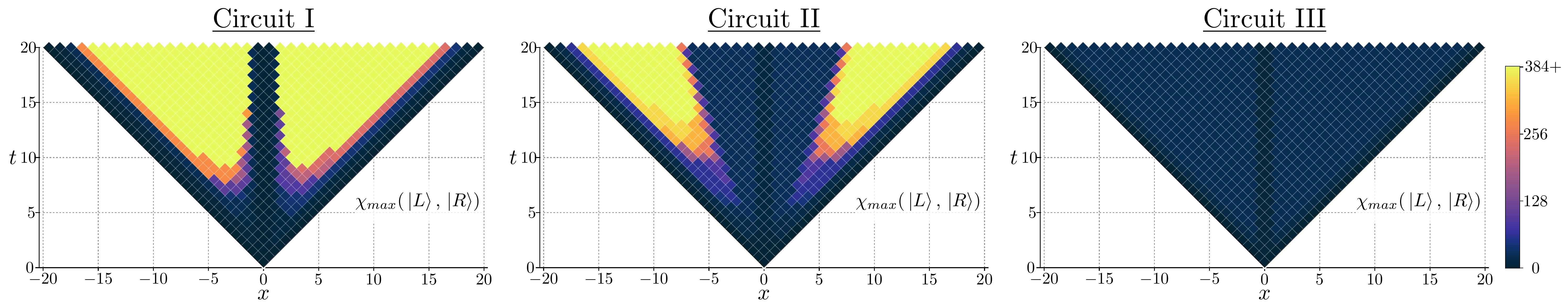}
    \caption{Maximum bond dimensions of both the (tilted) left and right influence matrices $\chi_{\text{max}}(\ket*{L_t},\ket*{R_t})$ encountered when computing each correlation function in Fig.~\ref{fig:correlations}. Regions that surpass the cutoff $\chi_c=384$ in the figure correspond to blank regions in Fig.~\ref{fig:correlations}. } 
    \label{fig:bonddims}
\end{figure*}

\textit{Correlation Functions.---}The various families of circuits characterised in this work display a diverse range of phenomenology. We illustrate this by numerically computing dynamical correlators Eq.~\eqref{eq:dynamicalcorr} in three circuits (I, II, and III) with different degrees of solvability, see Fig.~\ref{fig:correlations}. In the most general case, we observe nontrivial correlations along all rays $x=\zeta t$ within the lightcone ($|\zeta| \leq 1$), distinguishing them from the previously known families of solvable circuits (cf.~Refs.~\cite{bertini2019entanglement, piroli_exact_2020, giudice2021temporal, prosen2021many, jonay2021triunitary, claeys2024from, yu_hierarchical_2024, bertini2024exact, wang_2025_influence, rampp2025geometric, rampp2025solvable, rampp_2026_hierarchy, pickering2026asymptotically}), which only feature correlations along rays $\zeta$ in some discrete set (cf.~Circuit III). To obtain these data, we use an MPS-based algorithm to compute a \textit{tilted} influence matrix~\cite{foligno2023temporal}, whose computational cost is determined by the bond dimension $\chi$. See the End Matter for more details.

\textit{Conclusions.---}In this Letter we introduced a systematic, operator-trajectory inspired approach to generate quantum circuits with solvable, yet non-Markovian influence matrices.  We have shown that our approach admits an interpretation in terms of error-correction in the crossed channel and that yields families of circuits generating rich patterns of spatiotemporal correlations, where only those between points with zero separation are controlled by the influence matrices and therefore guaranteed to be solvable. However, we have also identified subfamilies featuring more broad forms of solvability at non-zero separation. 

Two natural questions for future research are to find an analytical characterisation for the full spatiotemporal correlation structure, i.e., find an efficient description of the `tilted' influence matrices mediating correlations between different spatial points, and analysing the quench dynamics produced by our family of circuits. We hope to report on these questions in future work.    

\emph{Acknowledgments.---} This work was supported by the Royal Society, through the University Research Fellowship No. 201101 (B.B.). M.M.~acknowledges support from Trinity College, Cambridge. B.K. acknowledges  support by the Novo Nordisk Foundation. 

{\it Note Added.---} While this manuscript was being finalised, we became aware of the related work~\cite{rampp_2026_hierarchy}. In our language this work constructs certain families of CDU2 circuits, whose correlation pattern is like that of Circuit III in Fig.~\ref{fig:correlations}.

\begin{center}
\begin{large}
\textbf{End Matter}
\end{large}
\end{center}

\begin{figure}[b]
    \includegraphics[width=1\columnwidth]{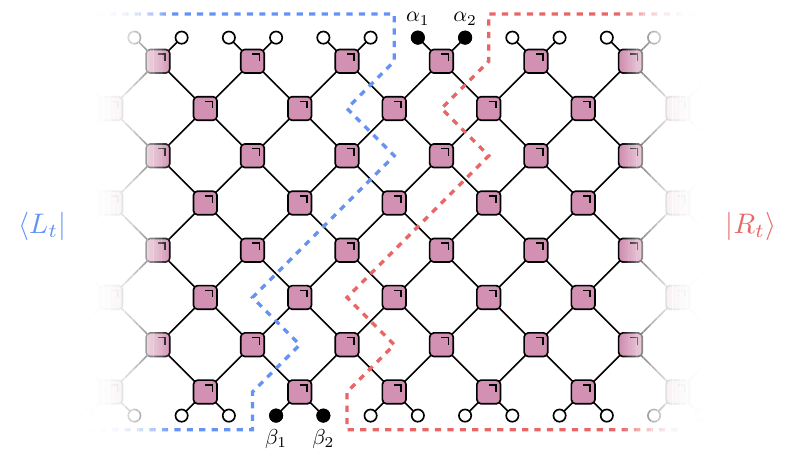}
    \caption{Tilted influence matrices characterising correlation functions with non-zero separation.}
    \label{fig:tilted_influence_matrices}
\end{figure}

\emph{Tilted influence matrices and correlations.---}To evaluate the correlation functions in Eq.~\eqref{eq:dynamicalcorr} for $x=\zeta t$ we use a diagonal contraction. Namely, we define \textit{tilted influence matrices}~\cite{foligno2023temporal}, where the vertical cut in Fig.~\ref{fig:influencematrix} is replaced by a space-time path from $(0,0)$ to $(x,t)$ (see~Fig.~\ref{fig:tilted_influence_matrices}), and seek to construct them as MPS states of bond dimension $\chi$ \textit{with no truncation}. When this can be achieved with a bond dimension lower than a defined cutoff $\chi_c$, we use it to compute the correlation function. If the bond-dimension required is larger than the cutoff $\chi_c$ we halt the computation. We do this for three different circuits (Circuit I--III), representing three different universality classes of our solutions (see Appendix~\ref{sec:tableau} for details) and fulfilling either CDU2 or CDU3. Although this implies that correlations for $x=0$ can be efficiently computed, the behaviour in other regions is highly dependent on the choice of gate. Our results are displayed in Fig.~\ref{fig:correlations} with a brief summary of the circuits  given in Table~\ref{tab:example-circuits}(see also Fig.~\ref{fig:bonddims} for a detailed account of the bond dimension used).

Circuit I shows the minimum solvability offered by the CDU$n$ conditions, where the bond-dimension required to capture any correlation with $\zeta\neq 0,1$ grows larger than the cutoff for sufficiently large $t$. In Circuit II we see a stronger form of solvability where there exists an inner light cone for which the tilted influence matrices have low bond-dimension. Circuit III is a completely solvable circuit where the bond-dimension is bounded for all points in space and time. The correlations can be described by contracting the tilted influence matrices analytically.

These qualitative differences can again be understood in terms of operator paths. In that language, tilted influence matrices mediate correlations if there exist paths that are able to enter and exit the environment,  see Fig.~\ref{fig:solvable_trajectories}b,c. For chiral fermionic Clifford circuits this cannot happen as the end-points of operator strings always travel at constant velocity: the only change in the operator growth occurs when a faster moving fermion located in the bulk of the string overtakes one of the end-points. This results in the operator string evolving as shown in Fig.~\ref{fig:solvable_trajectories}d where the end points draw out concave paths that, regardless of the inclination, cannot re-enter the system once they leave it. This means that the tilted influence matrices are also solvable. Deforming the circuit away from the free-fermionic Clifford point generically results in operator dynamics like that shown in Fig.~\ref{fig:solvable_trajectories}b,c, i.e., there will exist certain $\zeta$s for which operator paths can re-enter the system. However, certain longitudinal-field-like deformations do maintain the concave endpoint paths and thus the general solvability. More broadly, the phenomenology is differentiated by the existence of a lower bound to the operator end-point speeds, as in Circuit II, or not, as in Circuit I.

\begin{figure}
    \includegraphics[width=1\columnwidth]{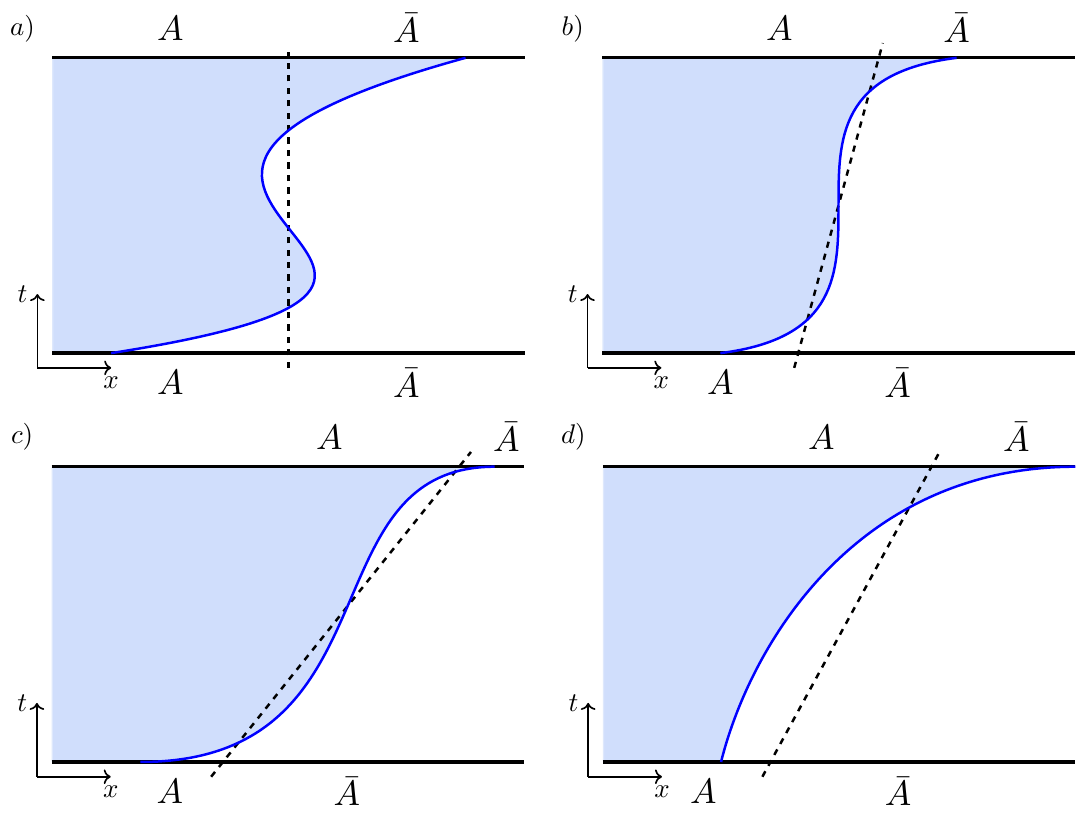}
    \caption{Examples of different operator trajectories and how they intersect with bipartitions at different tilt angles.}
    \label{fig:solvable_trajectories}
\end{figure}

\bibliography{ref}

\onecolumngrid
\appendix
\newpage

\section{Solvable influence matrices from operator trajectories}
\label{sec:operatortrajectories}

In the main text, we used the decomposition in Eq.~\eqref{eq:feynman path} to describe the Heisenberg evolution of a local operator, and used this to identify operator trajectories that lead to violation of the CDU conditions. Here, for completeness, we formulate a more precise relationship between operator trajectories and column dual-unitarity, which will allow us to derive a set of sufficient conditions for the CDU3 equations \eqref{eq:gdu3_columns} to be satisfied.

First, we remark that the influence matrix in Eq.~\eqref{eq:influencematrix} can be interpreted as a quantum channel (a completely positive and trace-preserving, or `CPTP', map) in the following sense. For each leg that is either dangling or contracted with a white circle tensor in Fig.~\ref{fig:influencematrix}a, we refer to it as `incoming' if it is connected to the bottom of some gate tensor, and `outgoing' if it is connected to the top of the gate tensor. We write $X$ for the subset of dangling legs that are incoming, and $Y$ for the subset of dangling legs that are outgoing; their cardinalities are $|X| = t$, $|Y| = t-1$. By designating $X$ as inputs and $Y$ as outputs, the operator $L_t$ can be reinterpreted as a linear map $\mathcal{N}_{Y \leftarrow X} : \mathcal{B}(\mathcal{H}_{X}) \rightarrow \mathcal{B}(\mathcal{H}_{Y})$, which (up to normalization) can be cast into the form
\begin{align}
    \mathcal{N}_{Y_t \leftarrow X_t}[O_{X_t}] = \tr_{B}\big[W(\pi_A \otimes O_{X_t})W^\dagger \big].
\end{align}
Here, $W$ is the unitary composed of all gates in the diagram in Fig.~\ref{fig:influencematrix}a, and $A$ ($B$) is the collection of incoming (outgoing) legs that are contracted with a circle tensor, and $\pi_A = I_A/d_A$ is the maximally mixed state on $A$, with $d_A = q^{|A|}$ the corresponding Hilbert space dimension. This is manifestly in the form of a CPTP map. By the same process, we can also reinterpret the left and right sides of Eq.~\eqref{eq:gdu3_columns} as linear maps with the same input and output spaces. We will write $\mathcal{N}^{(n)}_{Y \leftarrow X}$ for the channel in which $(n-1)$ columns of gates are kept. By Eq.~\eqref{eq:influencematrix}, we have $\mathcal{N}^{(n)}_{Y \leftarrow X} = \mathcal{N}_{Y \leftarrow X}$ for any $n \geq 2t+2$.

Let $\{P_\mu\}_{\mu=0}^{q^2-1}$ be a collection of operators on a single qudit, which are orthonormal $\tr[P_\mu P_\nu]/\tr[I] = \delta_{\mu \nu}$, with $P_{\mu = 0} = I$ (e.g.~Pauli operators if ${q = 2}$). By linearity, the map $\mathcal{N}_{Y \leftarrow X}^{(n)}$ (and by association, the corresponding influence matrix) is fully specified by the action of its adjoint $(\mathcal{N}^{(n)}_{Y \leftarrow X})^\dagger$ on strings of these operators over $Y$, namely $(\mathcal{N}^{(n)}_{Y \leftarrow X})^\dagger[P_{\nu_1} \otimes \cdots \otimes P_{\nu_{t-1}}]$, $\nu_\tau \in \{0, \ldots, q^2-1\}$ for $\tau = 1, \ldots t-1$. (We work in the Heisenberg picture here, where observables on the outputs are backward time-evolved by the adjoint channel.) Moreover, we can decompose the output in terms of strings of operators over $X$, so the collection of scalars
\begin{align}
    p^{(n)}\Big( (\nu_1 \ldots, \nu_{t-1}) \rightarrow (\lambda_1 \ldots \lambda_t)\Big) \coloneqq \tr\big[(P_{\lambda_1} \otimes \cdots \otimes P_{\lambda_t}) (\mathcal{N}^{(n)}_{Y \leftarrow X})^\dagger[P_{\nu_1} \otimes \cdots \otimes P_{\nu_{t-1}}]\big],
\end{align}
which can be seen as transition amplitudes in operator space. Due to the arrangement of gates in the definition of the influence matrix (see the diagram in Fig.~\ref{fig:influencematrix}a), it is helpful to work with the convention that $\nu_t = 0$.

For the full influence matrix $\mathcal{N}_{Y \leftarrow X} = \mathcal{N}_{Y \leftarrow X}^{(2t+2)}$, these amplitudes can also be decomposed in terms of operator trajectories, by analogy to Eq.~\eqref{eq:feynman path}. Specifically, we have
\begin{align}
    p^{(2t+2)}&\Big( (\nu_1 \ldots, \nu_{t-1}) \rightarrow (\lambda_1 \ldots  \lambda_t)\Big)\nonumber\\ &= \sum_{\substack{\mu \in \{0, \ldots, q^2-1\}^{2t\times(t+1) } \\  \vec{\mu}_t = \vec{0} }}  \delta_{\vec{\mu}_0 = 0} \times u\big( (\vec{\mu}_{t}, \nu_t) \rightarrow  (\vec{\mu}_{t-1}, \lambda_{t-1})\big) \cdots u\big( (\vec{\mu}_{1}, \nu_1) \rightarrow  (\vec{\mu}_{0}, \lambda_{0})\big).
    \label{eq:trajectory full}
\end{align}
Here, each $\mu \in \{0, \ldots, q^2-1\}^{2t\times(t+1) }$ represents a particular trajectory of operators $\mu(x, \tau)$, with $x \in \{x_0-1/2, x_0-1, \ldots, x_0-t\}$, and $\tau \in \{0, \ldots t\}$.  The shorthand $\vec{\mu}_\tau$ is used to denote all labels $\mu(x,\tau)$ for a fixed $\tau$, and the boundary conditions at the top and bottom of the circuit lead to the restriction $\vec{\mu}_0 = \vec{\mu_\tau} = \vec{0} \equiv (0 \ldots0)$. The function $u$ is given by
\begin{align}
    u\big((\vec{\mu}_{\tau}, \nu_{\tau})   \rightarrow  (\vec{\mu}_{\tau-1}, \lambda_{\tau-1})\big) \coloneqq \tr\big[(P_{\vec{\mu}_{\tau-1}} \otimes P_{\lambda_{\tau-1}}) \mathbb{U}^\dagger (P_{\vec{\mu}_\tau} \otimes P_{\nu_\tau})\mathbb{U}\big],
\end{align}
where $P_{\vec{\mu}_\tau} \coloneqq \bigotimes_{x\in [x_0-t,x_0-1/2]} P_{\mu(x,\tau)}$, and reflecting the arrangement of gates in the influence matrix, we assume the convention $\nu_t = 0$. 

With these coefficients, Equation \eqref{eq:trajectory full} constitutes a Feynman path representation for the evolution of operators within the spatial region $x \in [x_0-t,x_0-1/2]$, conditioned on a given set of `boundary conditions' $(\nu_\tau)_\tau$, $(\lambda_\tau)_\tau$. These boundary conditions specify the operators incoming and outgoing from the region $x \in [x_0-t,x_0-1/2]$ at its right boundary, and are therefore related to higher-order correlation functions of the form $\tr[\mathcal{O}_{x_0,t} \mathcal{O}_{x_0, t-1} \cdots \mathcal{O}_{x_0, 0}]$. Since we have to specify all these boundary operators on the whole temporal lattice, this  somewhat more cumbersome to specify, but we can still apply the intuition from the main text.

In particular, the advantage of this representation is that it becomes easy to see the effect of truncating the influence matrix to $n-1$ columns. Specifically, one only keeps operator trajectories in which $\mu(x,\tau) = 0$ for all $x \leq x_0 - n/2$,
\begin{align}
    p^{(n)}\Big((\lambda_1 \ldots \lambda_{t-1}\, 0) \leftarrow (\nu_1 \ldots, \nu_{t-1})\Big)= &\sum_{\substack{\mu \in \{0, \ldots, q^2-1\}^{2t\times(t+1) } \\  \vec{\mu}_t = \vec{0} }}   \delta_{\vec{\mu}_0 = 0}  \left( \prod_{x\in[x_0-t,x_0-n/2]} \prod_{\tau = 1}^{t-1} \delta_{\mu(x,\tau), 0} \right)\notag\\
    &\quad \times  u\big( (\vec{\mu}_{t}, \nu_t) \rightarrow  (\vec{\mu}_{t-1}, \lambda_{t-1})\big) \cdots u\big( (\vec{\mu}_{1}, \nu_1) \rightarrow  (\vec{\mu}_{0}, \lambda_{0})\big).
    \label{eq:trajectory full trunc}
\end{align}
Thus, if there are no operator trajectories with nonzero weight that 1) have nontrivial support in the region $x \leq x_0-n/2$ for some time $0 < \tau <t$, and 2) only have support on site $x_0$ at time $0$, then the influence matrix can be reduced to $(n-1)$ columns. This implies the corresponding CDU$n$ condition.

\section{Proof of Property \ref{prop:chiralfreefermCliff}}
\label{sec:property1}

As discussed in the main text, the CDU2 equations are solved in a Clifford circuit if we ensure that there exists a sufficient anticommutation structure between the left and right stabilisers of the columns, i.e. Eq.~\eqref{eq:unwanted}. Here we show that when the circuit is also free fermionic this structure is equivalent to imposing chirality on the fermion dynamics. Before beginning, we stress here that the statement that chiral fermion dynamics gives Markovian influence matrices is trivially shown by considering operator Feynman histories and is not the desired takeaway of Property \ref{prop:chiralfreefermCliff}. Rather, we want to demonstrate that the stabilizer machinery is equivalent for understanding this statement as this will be much more easily generalised beyond free fermionic circuits later.
We begin by expanding the notation of the main text that will be used in further appendices as well. Instead of priming indices in order to keep track of how they change when fermions pass through gates, we define the maps $P, Q, S, T$ such that
\be
\label{eq:app_index_maps_definition}
\begin{aligned}
\fineq[-.8ex][.8][1]{
    \roundgate[0][0][1][topright][bertinired]
    \smallcstate[-.35][-.35][][black]
    \node[scale=1.2] at (-1, -.35) {$\ell_{j}$};
}
= \sigma_j
\fineq[-.8ex][.8][1]{
    \roundgate[0][0][1][topright][bertinired]
    \smallcstate[-.35][.35][][black]
    \node[scale=1.2] at (-1, .35) {$\ell_{P(j)}$};
},
\quad j\in {\rm LL}, &\qquad 
\fineq[-.8ex][.8][1]{
    \roundgate[0][0][1][topright][bertinired]
    \smallcstate[-.35][-.35][][black]
    \node[scale=1.2] at (-.75, -.35) {$\mathcal P$};
    \smallcstate[.35][-.35][][black]
    \node[scale=1.2] at (.8, -.35) {$\ell_{k}$};
}
= \sigma_k
\fineq[-.8ex][.8][1]{
    \roundgate[0][0][1][topright][bertinired]
    \smallcstate[-.35][.35][][black]
    \node[scale=1.2] at (-.75, .35) {$\mathcal P$};
    \smallcstate[.35][.35][][black]
    \node[scale=1.2] at (1.1, .35) {$\ell_{Q(k)}$};
}, 
\quad k\in \rm{RR} \\
\fineq[-.8ex][.8][1]{
    \roundgate[0][0][1][topright][bertinired]
    \smallcstate[-.35][-.35][][black]
    \node[scale=1.2] at (-1, -.35) {$\ell_{j}$};
}
= \Tilde\sigma_j
\fineq[-.8ex][.8][1]{
    \roundgate[0][0][1][topright][bertinired]
    \smallcstate[.35][.35][][black]
    \node[scale=1.2] at (-.75, .35) {$\mathcal P$};
    \smallcstate[-.35][.35][][black]
    \node[scale=1.2] at (1.1, .35) {$\ell_{S(j)}$};
},
\quad j\in {\rm LR}, &\qquad 
\fineq[-.8ex][.8][1]{
    \roundgate[0][0][1][topright][bertinired]
    \smallcstate[-.35][-.35][][black]
    \node[scale=1.2] at (-.75, -.35) {$\mathcal P$};
    \smallcstate[.35][-.35][][black]
    \node[scale=1.2] at (.8, -.35) {$\ell_{k}$};
}
= \Tilde\sigma_k
\fineq[-.8ex][.8][1]{
    \roundgate[0][0][1][topright][bertinired]
    \smallcstate[-.35][.35][][black]
    \node[scale=1.2] at (-1, .35) {$\ell_{T(k)}$};
}, 
\quad k\in \rm{RL} \\
\end{aligned}
\ee
where $\mathcal P$ is the operator at position $x$ of $\mathcal{P}_x$, defined in Eq.~\eqref{eq:Px_definition}, and the $\sigma$'s are unimportant $\pm$ signs. By using $\mathcal{P}^2=I$ and that the gate commutes with the Jordan-Wigner string, i.e. $U (\mathcal{P}\otimes\mathcal{P})=(\mathcal{P}\otimes\mathcal{P}) U$, one may derive the equivalent expressions for the $r_j\equiv \mathcal P \ell_j$ operators.
We now explicitly state some of the objects relevant to the upcoming discussion
\be
\label{eq:app_columns}
\begin{aligned}
\fineq[-.8ex][.8][1]{
    \circletens[0][0][1][\col][2]
    \cstate[-.6][0]
    \cstate[0][-.6]
}
=
\fineq[-.8ex][.6][1]{
    \roundgate[0][0][1][topright][\col]
    \cstate[-.5][.5]
    \cstate[-.5][-.5]
    \roundgate[0][-2][1][topright][\col]
    \cstate[-.5][.5-2]
    \cstate[-.5][-.5-2]
    \node[scale=3,rotate=90] at (0, -4) {$\dots$};
    \roundgate[0][-6][1][topright][\col]
    \cstate[-.5][.5-6]
    \cstate[-.5][-.5-6]
    \draw[thick] (-.5,-7.5)--++(1,0);
    \cstate[-.5][-7.5]
}, \quad
\fineq[-.8ex][.8][1]{
    \circletens[0][0][1][\col][1]
    \cstate[0][.6]
}
=
\fineq[-.8ex][.6][1]{
    \draw[thick] (-.2,.5)--++(-.3,0);
    \draw[thick] (.2,.5)--++(.3,0);
    \cstate[-.2][.5]
    \cstate[.2][.5]
    \roundgate[0][-1][1][topright][\col]
    \roundgate[0][-3][1][topright][\col]
    \node[scale=3,rotate=90] at (0, -5) {$\dots$};
    \roundgate[0][-7][1][topright][\col]
},  \quad
\fineq[-.8ex][.8][1]{
    \circletens[0][0][1][\col][2]
    \cstate[-.6][0]
    \cstate[0][-.6]
    \circletens[1][0][1][\col][1]
    \cstate[1][.6]
}
=
\fineq[-.8ex][.6][1]{
    \roundgate[0][0][1][topright][\col]
    \cstate[-.5][.5]
    \cstate[-.5][-.5]
    \roundgate[0][-2][1][topright][\col]
    \cstate[-.5][.5-2]
    \cstate[-.5][-.5-2]
    \roundgate[0][-4][1][topright][\col]
    \cstate[-.5][.5-4]
    \cstate[-.5][-.5-4]
    \node[scale=3,rotate=90] at (.5, -6) {$\dots$};
    \roundgate[0][-8][1][topright][\col]
    \cstate[-.5][.5-8]
    \cstate[-.5][-.5-8]
    \draw[thick] (-.5,-9.5)--++(1,0);
    \cstate[-.5][-9.5]
    
    \draw[thick] (1-.2,.5)--++(-.3,0);
    \draw[thick] (1+.2,.5)--++(.3,0);
    \cstate[1-.2][.5]
    \cstate[1+.2][.5]
    \roundgate[1][-1][1][topright][\col]
    \roundgate[1][-3][1][topright][\col]
    \roundgate[1][-9][1][topright][\col]
}
=
\fineq[-.8ex][.6][1]{
    \roundgate[0][-2][1][topright][\col]
    \cstate[-.5][.5-2]
    \cstate[-.5][-.5-2]
    \roundgate[0][-4][1][topright][\col]
    \cstate[-.5][.5-4]
    \cstate[-.5][-.5-4]
    \node[scale=3,rotate=90] at (.5, -6) {$\dots$};
    \roundgate[0][-8][1][topright][\col]
    \cstate[-.5][.5-8]
    \cstate[-.5][-.5-8]

    \draw[thick] (.5,.5)--++(1,0);
    \cstate[.5][.5]
    \roundgate[1][-1][1][topright][\col]
    \roundgate[1][-3][1][topright][\col]
    \roundgate[1][-9][1][topright][\col]
    \cstate[1-.5][-.5]
    \cstate[1-.5][-9.5]
},
\quad
\ell_{\tau,j} =
\fineq[-.8ex][.6][1]{
    \draw[thick] (-.5, .5-2) --++ (1,0);
    \draw[thick] (-.5, -.5-2) --++ (1,0);
    \draw[thick] (-.5, .5-4) --++ (1,0);
    \draw[thick] (-.5, -.5-4) --++ (1,0);
    \draw[thick] (-.5, .5-8) --++ (1,0);
    \draw[thick] (-.5, -.5-8) --++ (1,0);
    \draw[thick] (-.5, .5) --++ (1,0);
    \draw[thick] (-.5, -.5) --++ (1,0);
    \draw[thick] (-.5, -9.5) --++ (1,0);
    \cstate[0][-.5-2][][]
    \cstate[0][.5-4][][]
    \node[scale=1.6] at (1.3, 1.5-4) {$\ell_{P(j)}$};
    \node[scale=1.6] at (1.1, .5-4) {$\ell_{j}$};
    \node[scale=3,rotate=90] at (0, -6) {$\dots$};
    \draw [decorate, thick, decoration = {brace}]   (1,.5)--++(0,-2);
    \node[scale=1.6, rotate=90] at (1.7, -.5) {$2(t-\tau)+1$};
    \draw [decorate, thick, decoration = {brace}]   (1,.5-5)--++(0,-5);
    \node[scale=1.6, rotate=90] at (1.7, 1-2-6) {$2(\tau-1)$};
}
\quad
r_{\tau,k} =
\fineq[-.8ex][.6][1]{
    \draw[thick] (-.5, .5-2) --++ (1,0);
    \draw[thick] (-.5, -.5-2) --++ (1,0);
    \draw[thick] (-.5, .5-4) --++ (1,0);
    \draw[thick] (-.5, -.5-4) --++ (1,0);
    \draw[thick] (-.5, .5-8) --++ (1,0);
    \draw[thick] (-.5, -.5-8) --++ (1,0);
    \draw[thick] (-.5, .5) --++ (1,0);
    \draw[thick] (-.5, -.5) --++ (1,0);
    \draw[thick] (-.5, -9.5) --++ (1,0);
    \cstate[0][.5-2][][]
    \cstate[0][-.5-2][][]
    \node[scale=1.6] at (1.3, .5-2) {$r_{Q(k)}$};
    \node[scale=1.6] at (1.1, -.5-2) {$r_{k}$};
    \node[scale=3,rotate=90] at (0, -6) {$\dots$};
    \draw [decorate, thick, decoration = {brace}]   (1,.5)--++(0,-1);
    \node[scale=1.6, rotate=90] at (1.7, .5-.5) {$2(t-\tau)$};
    \draw [decorate, thick, decoration = {brace}]   (1,.5-4)--++(0,-6);
    \node[scale=1.6, rotate=90] at (1.7, .5-1-6) {$2\tau - 1$};
}
\end{aligned}
\ee
We restate the expansion for a single gate and the full column
\be
\label{eq:app_explicit_columns}
\begin{aligned}
\fineq[-.8ex][.8][1]{
    \roundgate[-.5][0][1][topleft][bertiniblue]
    \roundgate[.5][0][1][topright][bertinired]
}
=
\prod_{k\in RR} \left(
\fineq[-0.8ex][0.7][1]{
    \draw(-.5, -.5)--++(1.0, 0);
    \draw(-.5,  .5)--++(1.0, 0);
}
+  \sigma_k
\fineq[-0.8ex][0.7][1]{
    \draw(-.5, -.5)--++(1.0, 0);
    \draw(-.5,  .5)--++(1.0, 0);
	\node[scale=1.3] at (-1.2,-.4)  {$r_k$};
	\node[scale=1.3] at (-1.2,.4)  {$r_{Q(k)}$};
    \cstate[0][-0.5][][black]
    \cstate[0][0.5][][black]
}\right) , \qquad
\fineq[-.8ex][.8][1]{
    \circletens[0][0][1][\col][2]
    \cstate[-.6][0]
    \cstate[0][-.6]
}  
=
\fineq[-.8ex][.8][1]{
    \draw[very thick] (-.5,0) -- (.5,0);
    \cstate[-.5][0]
}  
+
\sum_{j=1}^{N_R}
\fineq[-.8ex][.8][1]{
    \draw[very thick] (-.5,0) -- (.5,0);
    \cstate[-.5][0][][black]
	\node[scale=1.1] at (-1,-.05)  {$R_j$};
} 
\end{aligned}
\ee
The expansion of the gate is derived using the Clifford property to assert that the stabilisers fully characterise the non-zero elements of the channel.
From considering inserting the single-gate expansion into Eq.~\eqref{eq:app_columns}, one sees that the set of $R=\{R_j\}_{j=1,\dots,N_R}$ along with the identity and $L=\{L_j\}_{j=1,\dots,N_L}$ along with the identity form groups. As we discuss in the main text we want for each element of $R_i$ there to exist an $L_j$ that anticommutes $\{R_i, L_j\}=0$ as this is sufficient to show that the CDU2 conditions hold. We define the following generators of the groups
\be
\begin{aligned}
    \ell_{\tau, j} &= I\otimes \underbrace{I \otimes \dots \otimes I}_{2(t-\tau)} \otimes (\ell_{P(j)}\otimes \ell_{j}) \otimes\underbrace{ I \otimes \dots \otimes I}_{2(\tau-1)},&  &\tau = 1,\dots, t, \quad j\in {\rm LL},\\
    r_{\tau, k} &= \underbrace{I \otimes \dots \otimes I}_{2(t-\tau)} \otimes (r_{Q(k)}\otimes r_{k}) \otimes\underbrace{I\otimes \dots \otimes I}_{2(\tau-1)} \otimes I,&  &\tau = 1,\dots, t-1, \quad k\in {\rm RR}\label{{eq:l_r_stab}}
\end{aligned}
\ee
that are visualised above in Eq.~\eqref{eq:app_explicit_columns}. It is easy to see that these operators commute when their support does not overlap
\be
\label{eq:app_localmajorana_algebra}
\begin{aligned}
    [\ell_{\tau,j}, \ell_{\tau', j'}] = 0,& &[r_{\tau,k}, r_{\tau', k'}] = 0,& &
    [\ell_{\tau,j}, r_{\tau', k}] = 0 & &\text{if} \quad \tau' \neq \tau, \tau-1.
\end{aligned}
\ee
Instead, when $\tau' = \tau, \tau-1$ it is possible to have anticommutation between the left and right generators as they will overlap on a single leg. To understand this, we first remind the reader that the $\ell$'s and $r$'s are defined as the non-trivial heads of the Majorana fermions when the tail runs left or right respectively ($r_j\equiv\mathcal{P}\ell_j$). Therefore, they have the algebra
\be
    \{\ell_j, \ell_{j'}\} = 2\delta_{j,j'}, 
    \qquad \{r_k, r_{k'}\} = 2\delta_{k,k'},
    \qquad [\ell_j, r_{k}] = 2\delta_{j,k}\, \ell_j r_{k}
\ee
Now, we look at the generators $\tau' = \tau, \tau-1$ and find that the commutators are dependent on the $j,k$ indices as follows
\be
\label{eq:app_anticommuting_generators}
\begin{aligned}
   & \{\ell_{\tau,j}, r_{\tau, k}\} = 0 &&\text{if}\quad k=P(j) &&\text{and}& &[\ell_{\tau,j}, r_{\tau, k}] = 0& &\text{otherwise}, \\
   & \{\ell_{\tau,j}, r_{\tau-1, k}\} = 0&&\text{if}\quad j=Q(k) &&\text{and}& &[\ell_{\tau,j}, r_{\tau-1, k}] = 0 & &\text{otherwise},
\end{aligned}
\ee
which is most easily understood through the visual representation in Eq.~\eqref{eq:app_explicit_columns}. As we take these equations always for $j\in {\rm LL}$ and $k\in {\rm RR}$, the existence of an anticommuting $\ell$ above and below each $r$ is dependent on the structure of the sets ${\rm LL}$ and ${\rm RR}$ and therefore is a property of the fermion dynamics. However, in general we can make the following statement: \textit{for each right generator $r_{\tau,k}$ there exists at most one anticommuting $\ell_{\tau,j}$ and at most one anticommuting $\ell_{\tau+1,j}$, all other left generators commute}. 

We will now consider all the elements of $R$ as all possible products of the right generators. For each possible product we will attempt to find a corresponding product of left generators that anticommutes. When this is not possible we will identify the particular commutation structure responsible and add a restriction to the form of the gate, i.e.\ to the operator dynamics, that forbids it. We will demonstrate that the full set of restrictions amounts to imposing chirality on the fermion dynamics. 

We begin considering all products of generators involving a particular chosen $r_{1,k}$ which we simply denote as $\rho_1$. There either does or does not exist a corresponding anticommuting $\ell_{1,j}$, if there does we will denote it as $\lambda_1$ and if there does not then we will say $\lambda_1$ does not exist. We will graphically represent the two possibilities by drawing lines over the tensor network where these stabilisers connect as shown
\be
\begin{aligned}
\fineq[-.8ex][.7][1]{
    \roundgate[0][0][1][topright][\col]
    \roundgate[1][-1][1][topright][\col]
    \cstate[-.5][.5]
    \cstate[-.5][-.5]
    \cstate[.5][-1.5]
    \draw[very thick, color=red] (.5, -1.5) -- (1, -1) -- (0, 0) -- (.5, .5);
} \qquad \text{if $\lambda_1$ does exist and} \qquad 
\fineq[-.8ex][.7][1]{
    \roundgate[0][0][1][topright][\col]
    \roundgate[1][-1][1][topright][\col]
    \cstate[-.5][.5]
    \cstate[-.5][-.5]
    \cstate[.5][-1.5]
    \draw[very thick, color=red] (1.5, -1.5) -- (1, -1) -- (0, 0) -- (.5, .5);
} \qquad \text{if $\lambda_1$ does not exist.} 
\end{aligned}
\ee
In the first case, we have achieved our goal as any products of right generators that includes $\rho_1$ will anticommute with $\lambda_1$ as it commutes with all other generators. In the second case, however, we are not yet able to find an element of $L$ that anticommutes with each product of right generators involving $\rho_1$. We must now focus on this case. Having considered the left generator below $\rho_1$ we can now consider those above in $l_{2, j}$. Again there either does or does not exist one of these generators, dubbed $\lambda_2$, which anticommutes with $\rho_1$, for which we extend the previous diagram as
\be
\begin{aligned}
\fineq[-.8ex][.65][1]{
    \roundgate[0][0][1][topright][\col]
    \roundgate[1][-1][1][topright][\col]
    \roundgate[1][1][1][topright][\col]
    \cstate[-.5][.5]
    \cstate[-.5][-.5]
    \cstate[.5][-1.5]
    \draw[very thick, color=red] (1.5, -1.5) -- (1, -1) -- (0, 0) -- (1, 1) -- (.5, 1.5);
} \qquad \text{if $\lambda_2$ does exist and} \qquad
\fineq[-.8ex][.65][1]{
    \roundgate[0][0][1][topright][\col]
    \roundgate[1][-1][1][topright][\col]
    \roundgate[1][1][1][topright][\col]
    \cstate[-.5][.5]
    \cstate[-.5][-.5]
    \cstate[.5][-1.5]
    \draw[very thick, color=red] (1.5, -1.5) -- (1, -1) -- (0, 0) -- (1, 1) -- (1.5, 1.5);
} \qquad \text{if $\lambda_2$ does not exist.} 
\end{aligned}
\ee
The second diagram here now represents a complete failure to find any generators of $L$ that anticommute with our chosen $\rho_1$. This is not acceptable and we must impose a restriction on the gates to forbid it, i.e.\ we must always ensure that for each right generator we at least have one anticommuting left generator either above or below. As we continue to add restrictions like this it is simpler to just state which diagrams we forbid. Now we are left to consider the case where we have $\rho_1$ and $\lambda_2$ anticommuting. In this case we cannot immediately conclude that $\lambda_2$ anticommutes with all products of right generators including $\rho_1$ as there may exist another generator above ($\rho_2=r_{2,k}$ for some $k$) which also anticommutes with $\lambda_2$ such that all products including both $\rho_1$ and $\rho_2$ actually commute with $\lambda_2$. We represent the possibilities again by extending the diagram as
\be
\begin{aligned}
\fineq[-.8ex][.6][1]{
    \roundgate[0][0][1][topright][\col]
    \roundgate[1][-1][1][topright][\col]
    \roundgate[1][1][1][topright][\col]
    \roundgate[0][2][1][topright][\col]
    \cstate[-.5][2.5]
    \cstate[-.5][1.5]
    \cstate[-.5][.5]
    \cstate[-.5][-.5]
    \cstate[.5][-1.5]
    \draw[very thick, color=red] (1.5, -1.5) -- (1, -1) -- (0, 0) -- (1, 1) -- (0, 2) -- (.5, 2.5);
} \qquad \text{if $\rho_2$ does exist and} \qquad
\fineq[-.8ex][.6][1]{
    \roundgate[0][0][1][topright][\col]
    \roundgate[1][-1][1][topright][\col]
    \roundgate[1][1][1][topright][\col]
    \roundgate[0][2][1][topright][\col]
    \cstate[-.5][2.5]
    \cstate[-.5][1.5]
    \cstate[-.5][.5]
    \cstate[-.5][-.5]
    \cstate[.5][-1.5]
    \draw[very thick, color=red] (1.5, -1.5) -- (1, -1) -- (0, 0) -- (1, 1) -- (-.5, 2.5);
} \qquad \text{if $\rho_2$ does not exist.} 
\end{aligned}
\ee
Clearly the latter case ensures that all products involving $\rho_1$ anticommute with $\lambda_2$ but the former case does not. We thus now follow the former case and continue up the column again and consider $\lambda_3$. We will dispense with the discussion and just draw the diagrams with their meaning now understood. The next pair of diagrams are
\be
\begin{aligned}
\fineq[-.8ex][.6][1]{
    \roundgate[0][0][1][topright][\col]
    \roundgate[1][-1][1][topright][\col]
    \roundgate[1][1][1][topright][\col]
    \roundgate[0][2][1][topright][\col]
    \roundgate[1][3][1][topright][\col]
    \cstate[-.5][2.5]
    \cstate[-.5][1.5]
    \cstate[-.5][.5]
    \cstate[-.5][-.5]
    \cstate[.5][-1.5]
    \draw[very thick, color=red] (1.5, -1.5) -- (1, -1) -- (0, 0) -- (1, 1) -- (0, 2) -- (1, 3) -- (.5, 3.5);
} \qquad \text{and} \qquad
\fineq[-.8ex][.6][1]{
    \roundgate[0][0][1][topright][\col]
    \roundgate[1][-1][1][topright][\col]
    \roundgate[1][1][1][topright][\col]
    \roundgate[0][2][1][topright][\col]
    \roundgate[1][3][1][topright][\col]
    \cstate[-.5][2.5]
    \cstate[-.5][1.5]
    \cstate[-.5][.5]
    \cstate[-.5][-.5]
    \cstate[.5][-1.5]
    \draw[very thick, color=red] (1.5, -1.5) -- (1, -1) -- (0, 0) -- (1, 1) -- (0, 2) -- (1.5,3.5);
}. 
\end{aligned}
\ee
Clearly the second must be forbidden as it prevents the existence of a product of generators of $L$ anticommuting with $\rho_1\rho_2$. The first diagram is acceptable but again we cannot fully conclude without considering the next set of right generators above. At this points the process begins to repeat itself. At each new step we find ourselves forbidding a diagram where the line enters the network from the right, travels up and then exits to the right again. The process finally terminates when we reach the top of the columns and find ourselves considering 
\be
\begin{aligned}
\fineq[-.8ex][.6][1]{
    \roundgate[0][-2][1][topright][\col]
    \cstate[-.5][.5-2]
    \cstate[-.5][-.5-2]
    \roundgate[0][-4][1][topright][\col]
    \cstate[-.5][.5-4]
    \cstate[-.5][-.5-4]
    \node[scale=3,rotate=90] at (.5, -6) {$\dots$};
    \roundgate[0][-8][1][topright][\col]
    \cstate[-.5][.5-8]
    \cstate[-.5][-.5-8]

    \roundgate[1][-1][1][topright][\col]
    \roundgate[1][-3][1][topright][\col]
    \roundgate[1][-9][1][topright][\col]
    \cstate[1-.5][-.5]
    \cstate[1-.5][-9.5]
    
    \draw[very thick, color=red] (1.5, -9.5) --++ (-1.5,1.5) --++ (.5,.5);
    \draw[very thick, color=red] (.5, -4.5) --++ (-.5,.5) --++ (1,1) --++ (-1, 1) --++ (1, 1) --++ (-.5, .5);
} \qquad \text{and} \qquad
\fineq[-.8ex][.6][1]{
    \roundgate[0][-2][1][topright][\col]
    \cstate[-.5][.5-2]
    \cstate[-.5][-.5-2]
    \roundgate[0][-4][1][topright][\col]
    \cstate[-.5][.5-4]
    \cstate[-.5][-.5-4]
    \node[scale=3,rotate=90] at (.5, -6) {$\dots$};
    \roundgate[0][-8][1][topright][\col]
    \cstate[-.5][.5-8]
    \cstate[-.5][-.5-8]

    \roundgate[1][-1][1][topright][\col]
    \roundgate[1][-3][1][topright][\col]
    \roundgate[1][-9][1][topright][\col]
    \cstate[1-.5][-.5]
    \cstate[1-.5][-9.5]
    
    \draw[very thick, color=red] (1.5, -9.5) --++ (-1.5,1.5) --++ (.5,.5);
    \draw[very thick, color=red] (.5, -4.5) --++ (-.5,.5) --++ (1,1) --++ (-1, 1) --++ (1.5, 1.5);
}
\end{aligned}
\ee
we finally forbid the second diagram and note that the first diagram ensures that $\rho_1\rho_2\dots \rho_{t-1}$ anticommutes with $\lambda_2\dots \lambda_t$. In fact $\rho_1\rho_2\dots \rho_{t-1}$ anticommutes with just $\lambda_t$ which we would have recognised if we had started from the top of the column instead. Therefore, restricting ourselves to circuits where diagrams that feature the red line entering and exiting the influence matrix cannot occur, \textit{we find that for all products of generators including any of $r_{1,k}$ there exists an element of $L$ that anticommutes with the product}. As such we can now restrict ourselves to the set of elements of $R$ that are the identity on the lowest gate. The arguments now repeats itself for the remaining elements as we simply begin again for those that start on the second gate from the bottom and then the third and so on. In-fact we have already forbidden all the problematic diagrams that will show up there so no more restrictions are generated. We therefore find that \textit{restricting ourselves to circuits where diagrams that feature the red line entering and exiting the influence matrix cannot occur, for all elements of $R$ there exist an element of $L$ that anticommutes.} This is sufficient for the CDU2 conditions to hold.

All that we must do now is demonstrate that the class of free fermionic Clifford circuits which do not allow the forbidden diagrams are chiral. This is simple to do: we just need to recognise that the lines drawn in the diagrams are the possible world lines for the Majorana fermions to take through it. As we discussed in the main text, the left and right stabilisers encode fermions that do not hop when evolved by the gate. Furthermore, refering back to the conditions for anticommutation between the left and right generators in Eq.~\eqref{eq:app_anticommuting_generators}, we find anticommutation between a right generator and a left generator above only when $j=P(k)$ and between a right generator and a left generator below only when $k=P(j)$ both for $j\in {\rm LL}, k\in {\rm RR}$. We then have that evolving a Majorana operator corresponding to $j\in {\rm LL}$ yields
\be
\begin{aligned}
&\fineq[-0.8ex][.8][1]{
    \draw[thick] (.5,.5) --++(0,1);
    \draw[thick] (4+.5,.5) --++(0,-1);
    \cstate[4.5][-.5]
    \foreach \i in {1,...,2}{
        \roundgate[2*\i-1][0][1][topright][bertiniviolet]  
        \roundgate[2*\i][1][1][topright][bertiniviolet]  
    }
    \foreach \i in {1,...,1}{
        \cstate[2*\i-1-.5][-.5][][black]
        \cstate[2*\i-1+.5][-.5][][black]
        \node[scale=1] at (2*\i-1-.5, -.95)  {$\mathcal{P}$};
        \node[scale=1] at (2*\i-1+.5, -.95)  {$\mathcal{P}$};
    }
    \foreach \i in {2,...,2}{
        \cstate[2*\i-1-.5][-.5]
        \cstate[2*\i-1+.5][-.5]
    }
    \cstate[3-.5][-.5][][black]
    \node[scale=1.3] at (3-.5, -1.05)  {$\ell_j$};
}
=
\fineq[1.8ex][.8][1]{
    \draw[thick] (.5,.5) --++(0,1);
    \foreach \i in {1,...,2}{
        \roundgate[2*\i][1][1][topright][bertiniviolet]  
    }
    \foreach \i in {1,...,1}{
        \cstate[2*\i-1-.5][.5][][black]
        \cstate[2*\i-1+.5][.5][][black]
        \node[scale=1] at (2*\i-1-.5, 1-.95)  {$\mathcal{P}$};
        \node[scale=1] at (2*\i-1+.5, 1-.95)  {$\mathcal{P}$};
    }
    \foreach \i in {2,...,2}{
        \cstate[2*\i-1-.5][.5]
        \cstate[2*\i-1+.5][.5]
    }
    \cstate[3-.5][.5][][black]
    \node[scale=1.3] at (3, 1-1.05)  {$\ell_{P(j)}$};
    \cstate[4.5][.5]
} \\
&=
\fineq[-0.8ex][.8][1]{
    \foreach \i in {1,...,1}{
        \draw[thick] (2*\i-1-.5,.5) --++(0,1);
        \draw[thick] (2*\i-1+.5,.5) --++(0,1);
        \cstate[2*\i-1-.5][.5][][black]
        \cstate[2*\i-1+.5][.5][][black]
        \node[scale=1] at (2*\i-1-.5, 1-.95)  {$\mathcal{P}$};
        \node[scale=1] at (2*\i-1+.5, 1-.95)  {$\mathcal{P}$};
    }
    \foreach \i in {2,...,2}{
        \draw[thick] (2*\i-1-.5,.5) --++(0,1);
        \draw[thick] (2*\i-1+.5,.5) --++(0,1);
        \cstate[2*\i-1-.5][.5]
        \cstate[2*\i-1+.5][.5]
    }
    \cstate[3-.5][.5][][black]
    \node[scale=1.3] at (3, 1-1.05)  {$\ell_{Q(P(j))}$};
} \quad\text{if $P(j)\in {\rm RR}$}\quad \text{ and } \quad
\fineq[-0.8ex][.8][1]{
    \draw[thick] (.5,.5) --++(0,1);
    \foreach \i in {1,...,1}{
        \draw[thick] (2*\i-1-.5,.5) --++(0,1);
        \draw[thick] (2*\i-1+.5,.5) --++(0,1);
        \cstate[2*\i-1-.5][.5][][black]
        \cstate[2*\i-1+.5][.5][][black]
        \node[scale=1] at (2*\i-1-.5, 1-.95)  {$\mathcal{P}$};
    }
    \foreach \i in {2,...,2}{
        \draw[thick] (2*\i-1-.5,.5) --++(0,1);
        \draw[thick] (2*\i-1+.5,.5) --++(0,1);
        \cstate[2*\i-1-.5][.5]
        \cstate[2*\i-1+.5][.5]
    }
    \node[scale=1.3] at (2, 1-1.05)  {$\ell_{T(P(j))}$};
}\quad \text{otherwise},
\end{aligned}
\ee
where we remind the reader that the maps are defined in Eq.~\eqref{eq:app_index_maps_definition}. Therefore the condition for the fermion to remain static is the same for the next anticommuting generator to exist and the condition to hop is the same for the next anticommuting generator to not exist. It is easily seen that this matches the graphical notation we were using. This means the lines in the diagrams map directly to possible worldlines for the fermions. As such, all the forbidden diagrams represent a fermion hopping left and then hopping right again after some amount of time. If we forbid this from ever occurring, i.e.\ impose chirality such that any fermion that hops left may never hop right again. We find that the diagrams are forbidden. When one considers the CDU2 conditions for the right influence matrix the equivalent condition for right movers is found.

\section{Proof of Property \ref{prop:cdu2_general}}
\label{sec:property2}

Here we follow on from the discussion of the previous appendix and consider how we may deform the chiral free fermionic Clifford gates whilst preserving the CDU2 conditions. Specifically we consider defining the deformed gate $W$ that is related to the original Clifford gate $U$ by
\begin{equation}
    W = (u_1\otimes u_2)\,\, U \,\, (u_3\otimes u_4), \qquad u_1, u_2, u_3, u_4 \in {\rm U}(q).
\end{equation}
Perturbations of this form are particularly convenient to consider as they will never be able to alter the number of stabilisers of the gate. Their only action is to rotate the stabilisers such that we have
\be
\label{eq:app_rotated_stabilisers}
\begin{aligned}
&\fineq[-.8ex][.8][1]{
    \roundgate[0][0][1][topright][bertinired]
    \smallcstate[-.35][.35][][black]
    \node[scale=1.2] at (-1, .35) {$\ell_{P(j)}^{(2)}$};
    \smallcstate[-.35][-.35][][black]
    \node[scale=1.2] at (-1, -.35) {$\ell_j^{(1)}$};
}
\equiv
\fineq[-.8ex][.8][1]{
    \roundgate[0][0][1][topright][bertinired]
    \smallcstate[-.35][.35][][black]
    \node[scale=1.2] at (-1.4, .35) {$u_1\ell_{P(j)}u_1^\dag$};
    \smallcstate[-.35][-.35][][black]
    \node[scale=1.2] at (-1.4, -.35) {$u_3^\dag\ell_{j}u_3$};
}
= \sigma_j
\fineq[-.8ex][.8][1]{
    \roundgate[0][0][1][topright][bertinired]
},
& & j\in {\rm LL}, \\
& \fineq[-.8ex][.8][1]{
    \roundgate[0][0][1][topright][bertinired]
    \smallcstate[.35][.35][][black]
    \node[scale=1.2] at (1, .35) {$r_{Q(k)}^{(2)}$};
    \smallcstate[.35][-.35][][black]
    \node[scale=1.2] at (1, -.35) {$r_k^{(1)}$};
}\equiv
\fineq[-.8ex][.8][1]{
    \roundgate[0][0][1][topright][bertinired]
    \smallcstate[.35][.35][][black]
    \node[scale=1.2] at (1.6, .35) {$u_2 r_{Q(k)}u_2^\dag$};
    \smallcstate[.35][-.35][][black]
    \node[scale=1.2] at (1.6, -.35) {$u_4^\dag r_{k} u_4$};
}
= \sigma_k
\fineq[-.8ex][.8][1]{
    \roundgate[0][0][1][topright][bertinired]
},
&& k\in {\rm RR}.
\end{aligned}
\ee
This means we simply need to redefine the generators of the left and right stabiliser groups on the columns to 
\be
\begin{aligned}
    \ell_{\tau, j} &= I\otimes \underbrace{I\otimes \dots \otimes I}_{2(t-\tau)} \otimes (\ell_{P(j)}^{(2)}\otimes \ell_{j}^{(1)}) \otimes\underbrace{ I \otimes \dots \otimes I}_{2(\tau-1)},&  &\tau = 1,\dots, t, \quad j\in {\rm LL},\\
    r_{\tau, k} &= \underbrace{I \otimes \dots \otimes I}_{2(t-\tau)} \otimes (r_{Q(k)}^{(2)}\otimes r_{k}^{(1)}) \otimes\underbrace{I\otimes \dots \otimes I}_{2(\tau-1)} \otimes I,&  &\tau = 1,\dots, t-1, \quad k\in {\rm RR}\label{{eq:l_r_stab}}
\end{aligned}
\ee
As discussed in the proof of Property \ref{prop:chiralfreefermCliff}, for any free fermionic Clifford circuit the algebra between these generators is always commutation or anticommutation. 
Property \ref{prop:chiralfreefermCliff} then states that if the base gates are chiral then the algebra is restricted such that the CDU2 equations follow from it.
As such, \textit{if we deform a chiral free fermionic Clifford gate in the manner described above and maintain the algebra between the generators of the left and right stabilisers then the CDU2 equations are preserved}. 

The seemingly simplest way to do this is to simply construct the four unitaries such that the stabilisers are left invariant. This ends up covering more possible transformations than one might expect due to the gauge symmetry present in brickwork circuits where the bulk of the circuit is invariant under $U\rightarrow (v\otimes w) U (w^\dag \otimes v^\dag)$. We therefore want the following
\be
\begin{aligned}
    [u_1, \ell_{P(j)}] = [u_3, \ell_{j}] = 0,& \qquad j \in \rm LL \\
    [u_2, r_{Q(k)}] = [u_4, r_{k}] = 0,& \qquad k \in \rm RR
\end{aligned}
\ee
This can be achieved by constructing unitaries that live within algebras that we know commute with the stabilisers. For example, for $u_3$ we want to find a set of operators that commute with $\ell_{j},\,\, j\in \rm LL$ which is easily done, appealing to the mutual algebra of $\ell$'s and $r$'s (Eq.~\eqref{eq:app_localmajorana_algebra}), by selecting $r_k$ with $k$ from a set that is disjoint with $\rm LL$. This is clearly the set $\rm LR$, i.e. the set of fermions on the left legs of gates that hop to the right instead of staying still. Therefore, we generate an algebra using the set $\{r_k\}_{k\in\rm LR}$ and have $u_3$ be contained within it. We can repeat this game for the other three and get in total
\be
\begin{aligned}
    u_1& \in \mathcal A( \{r_{T(k)}\}_{k\in {\rm RL}}) \\
    u_2& \in \mathcal A( \{\ell_{S(j)}\}_{j\in {\rm LR}}) \\
    u_3& \in \mathcal A(  \{r_{k}\}_{k\in {\rm LR}}) \\
    u_4& \in \mathcal A(   \{\ell_{j}\}_{j\in {\rm RL}})
\end{aligned}
\ee
where $\mathcal A(\{O_n\}_{n=1}^N)$ is the algebra generated by the set of operators $\{O_n\}_{n=1}^N$.

\section{Tableau and Deformations for Example Circuits}
\label{sec:tableau}

Here we give the tableau for the three free fermionic Clifford circuits discussed in the main text. We do this by specifying the adjoint action of the gates upon the following generators of the Pauli group on $2n$ qubits
\be
\gamma_{2j-1} = \left(\prod_{k=1}^{j-1} X_k\right) Y_j, \qquad
\gamma_{2j} = \left(\prod_{k=1}^{j-1} X_k\right) Z_j
\ee

\begin{table}[h]
\centering
a)
\begin{tabular}{|c | c|} 
 \hline
 $\gamma_j$ & $U_{\rm I}\gamma_j U_{\rm I}^\dag$ \\ [0.5ex] 
 \hline\hline
 YIII & XXYI\\ 
 ZIII & XYII\\ 
 XYII & ZIII\\ 
 XZII & XXXZ\\
 XXYI & YIII\\
 XXZI & XXXY\\
 XXXY & XXZI\\
 XXXZ & XZII\\[1ex] 
 \hline
\end{tabular}
b)
\begin{tabular}{|c | c|} 
 \hline
 $\gamma_j$ & $U_{\rm II}\gamma_j U_{\rm II}^\dag$ \\ [0.5ex] 
 \hline\hline
 YIII & XXYI\\ 
 ZIII & ZIII\\ 
 XYII & XXXZ\\ 
 XZII & -XYII\\
 XXYI & XXZI\\
 XXZI & -YIII\\
 XXXY & XXXY\\
 XXXZ & XZII\\[1ex] 
 \hline
\end{tabular}
\quad c)
\begin{tabular}{|c | c|} 
 \hline
 $\gamma_j$ & $U_{\rm III}\gamma_j U_{\rm III}^\dag$ \\ [0.5ex] 
 \hline\hline
 YIIIII &-XXXYII\\ 
 ZIIIII & ZIIIII\\ 
 XYIIII & XZIIII\\ 
 XZIIII & XXXZII\\
 XXYIII & XYIIII\\
 XXZIII & XXXXXZ\\
 XXXYII & YIIIII\\
 XXXZII & XXXXZI\\
 XXXXYI & XXYIII\\
 XXXXZI & XXXXYI\\
 XXXXXY & XXXXXY\\
 XXXXXZ & XXZIII\\ [1ex] 
 \hline
\end{tabular}
\caption{Adjoint action of the different Clifford gates used for circuits I-III upon the local Majoranas. a) Table for circuit I, b) Table for circuit II, c) Table for circuit III.}
\label{table:tableau}
\end{table}

\noindent We now show the deformations of these gates used to produce Circuits I--III in Fig.~\ref{fig:correlations}. Circuit I is defined as   
\be
    W_{\rm I} = U_{\rm I} \, \exp(i(\theta_1 XIII + \theta_2 IIIX + \theta_3 IYII + \theta_4 IIYI)),
\ee
with $\theta_1=\theta_2=\pi/8$ and $\theta_3=\theta_4=\pi/10$. The base Clifford gate supports four bands of fermions with velocities $v\in\{-1,0,0,1\}$. The terms involving $X$ are of course quadratic in the Majoranas and can be seen to provide hybridisation between the $v=-1$ and first $v=0$ band and then between the $v=1$ and second $v=0$ bands. This creates different right-moving and left-moving bands supporting unbounded lower speeds. Finally the $Y$ terms are linear in the generators of the local algebras and therefore longitudinal field like terms which break the integrability of the circuit. It is important to note that this circuit lies outside the hypothesis of Property~\ref{prop:cdu2_general} as the quadratic Majorana terms involve those at velocity $v=0$ that do not move in the original Clifford circuit and therefore do not generically commute with all left and right stabilisers. However, due to the particular commutation structure that occurs for $v=0$ paths this ends up not being a problem. It is also interesting that the lack of a lower bound on the non-zero velocities of the fermions prevents us from writing down a local graphical condition that can be used to derive CDU2. 

Circuit II is defined as  
\be
    W_{\rm{II}} =  \exp(i(\theta_1 ZZII + \theta_2 IIZZ)) \, U_{\rm{II}} \, \exp(i\theta_3 ZZZZ),
\ee
with $\theta_1=\theta_2=\pi/8$ and $\theta_3=\pi/6$. The base Clifford gate supports four bands of fermions with velocities $v\in\{-1,-1/3,1/3,1\}$. The $ZZ$ rotations are quadratic forms of Majoranas that produce hybridisation between the left moving bands and right moving bands. The $ZZZZ$ term is then a four Majorana scattering term that promotes the circuit to CDU3 as described in Property \ref{prop:cdu3_general}.

Circuit III is defined as 
\be
    W_{\rm{III}} = \left( \exp(i\theta_1(\cos\phi_1 IIY + \sin\phi_1 ZXX))\otimes\exp(i\theta_2(\cos\phi_2 YII + \sin\phi_2 XXZ))\right) \, U_{\rm{III}},
\ee
with $\theta_1=\theta_2=\pi/(6\sqrt{2})$ and $\phi_1=\phi_2=\pi/4$. This gate fulfils Property \ref{prop:cdu2_general}. The base Clifford gate supports five bands of fermions with velocities $v\in\{-1,-1/3,0,1/3,1\}$. To interpret the single-site unitaries one may note that
\be
    \exp(i\theta(\cos\phi IIY + \sin\phi ZXX)) = \cos\theta III + i\sin\theta (\cos\phi IIY + \sin\phi ZXX), 
\ee
and similarly for the other qudit. Recognising that YII and ZXX are linear in the $\ell$'s and $r$'s we see that they are longitudinal-field-like terms. Specifically, they correspond to Majoranas with $|v|=1$. As such, they will inject and extract Majoranas with the maximum velocities to operator strings.


\begin{table*}[t]
    \centering
    \renewcommand{\arraystretch}{1.2}
    \setlength{\tabcolsep}{5pt}
    \begin{tabular}{c c c c c}
        \toprule
        Circuit
        &
        Clifford Velocities
        &
        Class
        &
        Deformation
        &
        Tilted solvability
        \\
        \midrule
        I
        &
        $\{-1,0,0,1\}$
        &
        CDU$_2$
        &
        Quadratic (dispersion) + linear (longitudinal-field-like)
        &
        $x=0$ only
        \\
        II
        &
        $\{-1,-1/3,1/3,1\}$
        &
        CDU$_3$
        &
        Quadratic (dispersion) + quartic scattering \eqref{eq:pert}
        &
        Inner cone
        \\
        III
        &
        $\{-1,-1/3,0,1/3,1\}$
        &
        CDU$_2$
        &
        Fine-tuned longitudinal-field-like
        &
        All tilts
        \\
        \bottomrule
    \end{tabular}
    \caption{
        Summary of the three example circuits shown in
        Fig.~\ref{fig:correlations}. The velocities refer to the
        Majorana bands of the underlying Clifford circuits.
    }
    \label{tab:example-circuits}
\end{table*}

\section{Proof of Property~\ref{prop:cdu3_general}}
\label{sec:property3}

As discussed in the main text, in order to solve the CDU3 conditions we require left stabilisers of the two column tensor that are not stabilisers of the single column. This is not possible for the chiral free fermionic circuits we have discussed so far. Instead of generalising the free fermionic discussion of Appendix \ref{sec:property1} to more columns, we will take a different route and introduce interactions to the circuits we have identified for CDU2. We will restrict ourselves to examples of circuits that are homogeneous, i.e.\ all gates are parametrised identically, and that have velocities with the following bounds
\be
\label{eq:app_velocity_bounds}
    v \in [-1, -v_0] \cup \{0\} \cup [v_0, 1],
\ee
where the non-zero bounds are tight in the sense that there really do exist chiral fermions propagating at $\pm1, \pm v_0$. For simplicity we will consider the fermions moving at velocity $v=1$ to be in minimal cycles. By this we mean that, defining the even and odd layers of the propagator as
\be
 \mathbb U = \mathbb{U}_o \mathbb{U}_e, 
 \qquad \mathbb{U}_o=\bigotimes_{x=1}^{L} U_{x-1/2,x}, 
 \qquad \mathbb{U}_e= \bigotimes_{x=0}^{L-1} U_{x,x+1/2}
\ee
we can find Majorana operators that, for integer $x$, evolve as follows
\be
\begin{aligned}
&\mathbb U_e \gamma_x^{(1)} \mathbb U_e ^\dag = \gamma_{x+1/2}^{(1)}, & & \mathbb U_e \gamma_{x+1/2}^{(1)} \mathbb U_e ^\dag = \gamma_{x}^{(1)}, \\
&\mathbb U_o \gamma_{x+1/2}^{(1)} \mathbb U_o ^\dag = \gamma_{x+1}^{(1)}, & & \mathbb U_o \gamma_{x}^{(1)} \mathbb U_o ^\dag = \gamma_{x-1/2}^{(1)}. \\
\end{aligned}
\ee
We will then take $v_0 = 1/T$ for odd integer $T$ as this is easily realisable on the circuit geometry. In particular we again formulate the velocities with cycles as simple as possible such that
\be
\begin{aligned}
&\mathbb U_e \gamma_{1,x}^{(v_0)} \mathbb U_e ^\dag = \gamma_{2,x}^{(v_0)}, & & \mathbb U_e \gamma_{1,x+1/2}^{(v_0)} \mathbb U_e ^\dag = \gamma_{2,x+1/2}^{(v_0)} \\
&\mathbb U_o \gamma_{2,x}^{(v_0)} \mathbb U_o ^\dag = \gamma_{3,x}^{(v_0)}, & & \mathbb U_o \gamma_{2,x+1/2}^{(v_0)} \mathbb U_o ^\dag = \gamma_{3,x+1/2}^{(v_0)} \\
&& \dots &\\
&\mathbb U_e \gamma_{T,x}^{(v_0)} \mathbb U_e ^\dag = \gamma_{1,x+1/2}^{(v_0)}, & & \mathbb U_e \gamma_{T,x+1/2}^{(v_0)} \mathbb U_e ^\dag = \gamma_{1,x}^{(v_0)} \\
\end{aligned}
\ee
where the cycle now repeats by noticing that from homogeneity we have that $\mathbb U_o$ is $\mathbb U_e$ shifted by one site. There is an important consequence to the velocity bounds Eq.~\eqref{eq:app_velocity_bounds} from the perspective of Property~\ref{prop:chiralfreefermCliff} and its proof. As fermions can only enter and remain within two columns for a maximum time $T/2$ we find that anticommuting left stabilisers can be constructed for all elements of $R$ involving a given right generator by using left generators that are close to the right generators involved. In this case, the locality implies the existence of a local condition upon the gates similar to DU and DU2. For various choices of $v_0$ one has
\be
\label{eq:app_local_graphical_conditions_CDU2}
\begin{aligned}
\fineq[-.8ex][.6][1]{
    \roundgate[0][0][1][topright][\col]
    \roundgate[1][1][1][topright][\col]
    \roundgate[0][2][1][topright][\col]
    \roundgate[1][3][1][topright][\col]
    \cstate[.5][3.5]
    \cstate[-.5][2.5]
    \cstate[-.5][1.5]
    \cstate[-.5][.5]
    \cstate[-.5][-.5]
} = 
\fineq[-.8ex][.6][1]{
    \roundgate[0][0][1][topright][\col]
    \roundgate[1][1][1][topright][\col]
    \roundgate[1][3][1][topright][\col]
    \cstate[.5][3.5]
    \cstate[.5][2.5]
    \cstate[.5][1.5]
    \cstate[-.5][.5]
    \cstate[-.5][-.5]
}, \quad \text{for } v_0=\frac{1}{3}, \qquad
\fineq[-.8ex][.6][1]{
    \roundgate[0][0][1][topright][\col]
    \roundgate[1][1][1][topright][\col]
    \roundgate[0][2][1][topright][\col]
    \roundgate[1][3][1][topright][\col]
    \roundgate[0][4][1][topright][\col]
    \roundgate[1][5][1][topright][\col]
    \cstate[.5][5.5]
    \cstate[-.5][4.5]
    \cstate[-.5][3.5]
    \cstate[-.5][2.5]
    \cstate[-.5][1.5]
    \cstate[-.5][.5]
    \cstate[-.5][-.5]
} = 
\fineq[-.8ex][.6][1]{
    \roundgate[0][0][1][topright][\col]
    \roundgate[1][1][1][topright][\col]
    \roundgate[0][2][1][topright][\col]
    \roundgate[1][3][1][topright][\col]
    \roundgate[1][5][1][topright][\col]
    \cstate[.5][5.5]
    \cstate[.5][4.5]
    \cstate[.5][3.5]
    \cstate[-.5][2.5]
    \cstate[-.5][1.5]
    \cstate[-.5][.5]
    \cstate[-.5][-.5]
},\quad \text{for } v_0=\frac{1}{5} \text{ and so on ...}
\end{aligned}
\ee
As one can see, decreasing $v_0$ and thus increasing $T$ has the effect of increasing the height of the graphical condition. This can be seen simply by drawing the fermion paths through the diagram as shown in Appendix \ref{sec:property1}. There are in fact stronger graphical conditions that can be derived for free fermionic Clifford gates with these velocity restrictions but these will not be relevant to the deformed gates we consider now.

Consider adding the perturbations described in the main text, i.e.
\be
    W_{x,x+1/2} = U_{x,x+1/2} \exp(i\theta \gamma_x^{(1)}\gamma_{1,x}^{(v_0)}\gamma_{1,x+1/2}^{(v_0)}\gamma_{x+1/2}^{(1)}). 
\ee
The first effect of adding this perturbation to the gate is that we lose a single right and left stabiliser group generator corresponding to $\gamma_{1,x+1/2}^{(v_0)}$ and $\gamma_{1,x}^{(v_0)}$ respectively. This happens simply because these fermions when evolved are, with partial amplitude, scattered into non-local products of fermions and therefore do not remain completely localised on either side of the gate. We will denote by $\rm LL'$ and $\rm RR'$ the set of remaining stabilizer generators and by $\alpha = {\rm LL} \setminus \rm LL'$ and $\beta = \rm RR \setminus \rm RR'$ the removed stabiliser indices such that
\be
\begin{aligned}
\sigma_j  \fineq[-.8ex][.8][1]{
    \roundgate[0][0][1][topright][bertiniredstronger]
    \smallcstate[-.35][.35][][black]
    \node[scale=1.2] at (-1, .35) {$\ell_{P(j)}$};
    \smallcstate[-.35][-.35][][black]
    \node[scale=1.2] at (-1, -.35) {$\ell_{j}$};
}
&= 
\fineq[-.8ex][.8][1]{
    \roundgate[0][0][1][topright][bertiniredstronger]
},
\quad j\in {\rm LL'}, \\
\sigma_j  \fineq[-.8ex][.8][1]{
    \roundgate[0][0][1][topright][bertiniredstronger]
    \smallcstate[.35][.35][][black]
    \node[scale=1.2] at (1, .35) {$r_{Q(j)}$};
    \smallcstate[.35][-.35][][black]
    \node[scale=1.2] at (.8, -.35) {$r_{j}$};
}
&= 
\fineq[-.8ex][.8][1]{
    \roundgate[0][0][1][topright][bertiniredstronger]
},
\quad j\in {\rm RR'}, 
\end{aligned}
\ee
where the darker gates represent the perturbed gates $W_{x,x+1/2}$. One can also show that the operator expansion is modified to
\be
\begin{aligned}
\fineq[-.8ex][.8][1]{
    \roundgate[-1][0][1][topleft][bertiniblue]
    \roundgate[0][0][1][topright][bertinired]
}
=
\prod_{j\in \rm RR'} \left(
\fineq[-0.8ex][0.7][1]{
    \draw(-.5, -.5)--++(1.0, 0);
    \draw(-.5,  .5)--++(1.0, 0);
}
+  \sigma_j
\fineq[-0.8ex][0.7][1]{
    \draw(-.5, -.5)--++(1.0, 0);
    \draw(-.5,  .5)--++(1.0, 0);
	\node[scale=1.3] at (-1.4,-.4)  {$r_{j}$};
	\node[scale=1.3] at (-1.4,.4)  {$r_{Q(j)}$};
    \cstate[0][-0.5][][black]
    \cstate[0][0.5][][black]
}\right)
\left(
\fineq[-0.8ex][0.7][1]{
    \draw(-.5, -.5)--++(1.0, 0);
    \draw(-.5,  .5)--++(1.0, 0);
}
+  \sigma_\beta\cos\theta
\fineq[-0.8ex][0.7][1]{
    \draw(-.5, -.5)--++(1.0, 0);
    \draw(-.5,  .5)--++(1.0, 0);
	\node[scale=1.3] at (-1.4,-.4)  {$r_\beta$};
	\node[scale=1.3] at (-1.4,.4)  {$r_{Q(\beta)}$};
    \cstate[0][-0.5][][black]
    \cstate[0][0.5][][black]
}\right).
\end{aligned}
\ee
The operators present in this expansion are identical to the original gate's expansion in Eq.~\eqref{eq:app_explicit_columns} but that the elements involving the removed stabiliser now have a coefficient of $\cos\theta$. As such, the total set of operators we want to vanish remains the same with some simply having a modified coefficient. The real problem from the perspective of satisfying CDU2 is that having removed a generator of the left stabiliser group of the gate and therefore many generators of the column's group, we will now have elements of $R$ that are not accounted for by an anticommuting left stabiliser. As such, the conditions in Eq.~\eqref{eq:app_local_graphical_conditions_CDU2} will not hold. As the stabilisers we have removed correspond to a fermion at the beggining of its cycle, i.e.\ it has just hopped to the left or right and now will stay in place for $T-1$ layers of gates before hopping again, we can easily identify the problematic paths. Drawing them on top of the diagrams we have 
\be
\label{eq:app_local_graphical_conditions_CDU2}
\begin{aligned}
\fineq[-.8ex][.6][1]{
    \roundgate[0][0][1][topright][\col]
    \roundgate[1][1][1][topright][\col]
    \roundgate[0][2][1][topright][\col]
    \roundgate[1][3][1][topright][\col]
    \cstate[.5][3.5]
    \cstate[-.5][2.5]
    \cstate[-.5][1.5]
    \cstate[-.5][.5]
    \cstate[-.5][-.5]
    \draw[very thick, color=green] (-.5,-.5) --++ (1.5,1.5) --++ (-1,1) --++ (1,1) --++ (.5, .5);
    \draw[very thick, color=red] (1-.5,1-.5) --++ (.5,.5) --++ (-.5,.5);
} , \qquad
\fineq[-.8ex][.6][1]{
    \roundgate[0][0][1][topright][\col]
    \roundgate[1][1][1][topright][\col]
    \roundgate[0][2][1][topright][\col]
    \roundgate[1][3][1][topright][\col]
    \roundgate[0][4][1][topright][\col]
    \roundgate[1][5][1][topright][\col]
    \cstate[.5][5.5]
    \cstate[-.5][4.5]
    \cstate[-.5][3.5]
    \cstate[-.5][2.5]
    \cstate[-.5][1.5]
    \cstate[-.5][.5]
    \cstate[-.5][-.5]
    \draw[very thick, color=green] (-.5,-.5) --++ (1.5,1.5) --++ (-1,1) --++ (1,1) --++ (-1, 1) --++ (1, 1) --++ (.5, .5);
    \draw[very thick, color=red] (1-.5,1-.5) --++ (.5,.5) --++ (-.5,.5);
}, \qquad
\dots
\end{aligned}
\ee
where the red part of the line indicates the stabiliser which is missing. The fact that there is only one possible path reliant on the missing stabiliser tells us that for each generator we have removed from the full column by adding the perturbation, there is only a single combination of generators of $R$ that is unaccounted for, i.e., 
\be
\label{eq:app_problematic_R}
    R_j =  I\otimes(r_{\alpha_{T-1}} \otimes r_{\alpha_{T-2}}) \otimes\dots\otimes (r_{\alpha_2} \otimes r_{\alpha_1})
\ee
where we have defined
\be
    \alpha_i =
    \begin{cases}
        P(\, (Q\cdot P)^{i-1} (\alpha)\,), &i \text{ odd}\\
        (Q\cdot P)^{i-1} (\alpha), &i \text{ even}\\
    \end{cases}
\ee
with $(Q\cdot P)(j)\equiv Q(P(j))$. Now, we will show that for this operator we can find a left stabiliser of the two columns that anticommutes.

We begin by deriving the identity eluded to in the main text. Despite the added scattering terms removing a stabiliser from the gate, one can show that when an additional column is added this stabiliser is recovered by deforming the scattering term through the network. We begin with
\be
    \exp(i\theta \gamma_x^{(1)}\gamma_{1,x}^{(v_0)}\gamma_{1,x+1/2}^{(v_0)}\gamma_{x+1/2}^{(1)}) =
     I^{\otimes 2(x-1)} \otimes (\cos\theta\,\, I\otimes I + i \sin\theta\,\, \ell_{j} \ell_{\alpha} \otimes r_{\beta} r_{k}) \otimes I^{\otimes 2(L-x)}
\ee
where the labels $j,k$ correspond respectively to the Majoranas $\gamma_x^{(1)}, \gamma_{x+1/2}^{(1)}$ which have had their labels suppressed in favour of their velocity index. Now consider deforming the non-identity term , i.e. for $T=5$ we have
\be
\label{eq:app_majorana_deformation}
\begin{aligned}
\fineq[-.8ex][.6][1]{
    \roundgate[0][0][1][topright][bertinired]
    \roundgate[1][1][1][topright][bertiniredstronger]
    \roundgate[0][2][1][topright][bertiniredstronger]
    \roundgate[1][3][1][topright][bertiniredstronger]
    \roundgate[0][4][1][topright][bertiniredstronger]
    \smallcstate[-.35][-.35][][]
    \smallcstate[.35][-.35][][]
    \draw [decorate, thick, decoration = {brace}]   (-1,-.5)--(-1,4.5);
    \node[scale=1.3] at (-2.2,2) {$T \text{ gates}$};
    \node[scale=1.3] at (-1.2,-.9) {$\ell_{j} \ell_{\alpha}$};
    \node[scale=1.3] at (1.3,-.9) {$r_{\beta} r_{k}$};
} \sim
\fineq[-.8ex][.6][1]{
    \roundgate[0][0][1][topright][bertinired]
    \roundgate[1][1][1][topright][bertiniredstronger]
    \roundgate[0][2][1][topright][bertiniredstronger]
    \roundgate[1][3][1][topright][bertiniredstronger]
    \roundgate[0][4][1][topright][bertiniredstronger]
    \smallcstate[-.35][-.35][][]
    \smallcstate[-.35][.35][][]
    \smallcstate[.5][.5][][]
    \node[scale=1.3] at (-1.2,-.9) {$\ell_{j} \ell_{\alpha}$};
    \node[scale=1.3] at (-1.1,.5) {$r_{k}$};
    \node[scale=1.3] at (1.1,.1) {$\ell_{\beta_1}$};
}  \sim
\fineq[-.8ex][.6][1]{
    \roundgate[0][0][1][topright][bertinired]
    \roundgate[1][1][1][topright][bertiniredstronger]
    \roundgate[0][2][1][topright][bertiniredstronger]
    \roundgate[1][3][1][topright][bertiniredstronger]
    \roundgate[0][4][1][topright][bertiniredstronger]
    \smallcstate[-.35][-.35][][]
    \smallcstate[-.35][.35][][]
    \smallcstate[.5][1.5][][]
    \node[scale=1.3] at (-1.2,-.9) {$\ell_{j} \ell_{\alpha}$};
    \node[scale=1.3] at (-1.1,.5) {$r_{k}$};
    \node[scale=1.3] at (1.2,2) {$\ell_{\beta_2}$};
} \sim
\fineq[-.8ex][.6][1]{
    \roundgate[0][0][1][topright][bertinired]
    \roundgate[1][1][1][topright][bertiniredstronger]
    \roundgate[0][2][1][topright][bertiniredstronger]
    \roundgate[1][3][1][topright][bertiniredstronger]
    \roundgate[0][4][1][topright][bertiniredstronger]
    \smallcstate[-.35][-.35][][]
    \smallcstate[-.35][.35][][]
    \smallcstate[-.35][2-.35][][]
    \smallcstate[-.35][2.35][][]
    \smallcstate[.35][2.35][][]
    \node[scale=1.3] at (-1.2,-.9) {$\ell_{j} \ell_{\alpha}$};
    \node[scale=1.3] at (-1.1,.5) {$r_{k}$};
    \node[scale=1.3] at (-1,2-.35) {$\mathcal P$};
    \node[scale=1.3] at (-1,2.35) {$\mathcal P$};
    \node[scale=1.3] at (1.2,2) {$\ell_{\beta_3}$};
} \sim
\fineq[-.8ex][.6][1]{
    \roundgate[0][0][1][topright][bertinired]
    \roundgate[1][1][1][topright][bertiniredstronger]
    \roundgate[0][2][1][topright][bertiniredstronger]
    \roundgate[1][3][1][topright][bertiniredstronger]
    \roundgate[0][4][1][topright][bertiniredstronger]
    \smallcstate[-.35][-.35][][]
    \smallcstate[-.35][.35][][]
    \smallcstate[-.35][2-.35][][]
    \smallcstate[-.35][2.35][][]
    \smallcstate[.5][3.5][][]
    \node[scale=1.3] at (-1.2,-.9) {$\ell_{j} \ell_{\alpha}$};
    \node[scale=1.3] at (-1.1,.5) {$r_{k}$};
    \node[scale=1.3] at (-1,2-.35) {$\mathcal P$};
    \node[scale=1.3] at (-1,2.35) {$\mathcal P$};
    \node[scale=1.3] at (1.2,4) {$\ell_{\beta_4}$};
} \sim
\fineq[-.8ex][.6][1]{
    \roundgate[0][0][1][topright][bertinired]
    \roundgate[1][1][1][topright][bertiniredstronger]
    \roundgate[0][2][1][topright][bertiniredstronger]
    \roundgate[1][3][1][topright][bertiniredstronger]
    \roundgate[0][4][1][topright][bertiniredstronger]
    \smallcstate[-.35][-.35][][]
    \smallcstate[-.35][.35][][]
    \smallcstate[-.35][2-.35][][]
    \smallcstate[-.35][2.35][][]
    \smallcstate[-.35][4-.35][][]
    \smallcstate[-.35][4.35][][]
    \node[scale=1.3] at (-1.2,-.9) {$\ell_{j} \ell_{\alpha}$};
    \node[scale=1.3] at (-1.1,.5) {$r_{k}$};
    \node[scale=1.3] at (-1,2-.35) {$\mathcal P$};
    \node[scale=1.3] at (-1,2.35) {$\mathcal P$};
    \node[scale=1.3] at (-1.2,4-.35) {$\mathcal P$};
    \node[scale=1.3] at (-1.2,4.35) {$\ell_{\beta}$};
} 
\end{aligned}
\ee
where we are using $\sim$ to represent equality up to potential minus signs that arise due to the choice of underlying Clifford gate and we define
\be
    \beta_i =
    \begin{cases}
        Q(\, (P\cdot Q)^{i-1} (\beta)\,), &i \text{ odd}\\
        (P\cdot Q)^{i-1} (\beta), &i \text{ even}\\
    \end{cases}.
\ee
For our purposes, it is enough to simply understand these relations as literally propagating the fermions making up the interaction as they would have if evolved in the Sch\"odinger picture, but stopping their evolution when they exit the two columns. As such, if we represent the original deformation with a line connecting the legs it acts on, we have the following type of identity
\be
\label{eq:app_clifford+scattering_deformation}
\begin{aligned}
\fineq[-.8ex][.7][1]{
    \roundgatewithline[0][0][bertinired]
    \roundgatewithline[1][1][bertinired]
    \roundgatewithline[0][2][bertinired]
    \roundgatewithline[1][3][bertinired]
    \roundgatewithline[0][4][bertinired]
    \draw[thick] (-.5,-.5) --++(-.5,0);
    \draw[thick] (-.5,1-.5) --++(-.5,0);
    \draw[thick] (-.5,2-.5) --++(-.5,0);
    \draw[thick] (-.5,3-.5) --++(-.5,0);
    \draw[thick] (-.5,4-.5) --++(-.5,0);
    \draw[thick] (-.5,5-.5) --++(-.5,0);
}=
\fineq[-.8ex][.7][1]{
    \roundgate[0][0][1][topright][bertinired]
    \roundgatewithline[1][1][bertinired]
    \roundgatewithline[0][2][bertinired]
    \roundgatewithline[1][3][bertinired]
    \roundgatewithline[0][4][bertinired]
    \draw[thick] (-.5,-.5) --++(-.5,0);
    \draw[thick] (-.5,1-.5) --++(-.5,0);
    \draw[thick] (-.5,2-.5) --++(-.5,0);
    \draw[thick] (-.5,3-.5) --++(-.5,0);
    \draw[thick] (-.5,4-.5) --++(-.5,0);
    \draw[thick] (-.5,5-.5) --++(-.5,0);
    \draw[very thick, color=gray] (-.75, -.5) --++ (0, 5);
    \draw[thick, fill=black] (-.75, -.5) circle (1pt); 
    \draw[thick, fill=black] (-.75, 1-.5) circle (1pt); 
    \draw[thick, fill=black] (-.75, 2-.5) circle (1pt); 
    \draw[thick, fill=black] (-.75, 3-.5) circle (1pt); 
    \draw[thick, fill=black] (-.75, 4-.5) circle (1pt); 
    \draw[thick, fill=black] (-.75, 5-.5) circle (1pt); 
    \node[scale=1.5] at (-.75,5) {$V$};
    
}
\end{aligned}
\ee
where the unitary applied to the side legs $V$ takes the form
\be
    V = e^{i\theta O}, \qquad O = (\ell_\beta \otimes \mathcal P) \otimes (\mathcal P \otimes \mathcal P)^{\otimes (T-3)/2} \otimes (r_{k} \otimes \ell_{j}\ell_{\alpha}). 
\ee
Note that by removing the scattering gate from underneath the lowest Clifford gate, we have essentially unlocked the previously removed stabiliser of that gate. The only caveat is that now with the unitary connecting many sites in time, this stabiliser will now grow across all these sites. That is to say we have the additional stabiliser
\be
\begin{aligned}
\sigma_j
\fineq[-.8ex][.7][1]{
    \roundgate[0][0][1][topright][bertiniredstronger]
    \roundgate[1][1][1][topright][bertiniredstronger]
    \roundgate[0][2][1][topright][bertiniredstronger]
    \roundgate[1][3][1][topright][bertiniredstronger]
    \roundgate[0][4][1][topright][bertiniredstronger]
    \draw[thick] (-.5,-.5) --++(-.5,0);
    \draw[thick] (-.5,1-.5) --++(-.5,0);
    \draw[thick] (-.5,2-.5) --++(-.5,0);
    \draw[thick] (-.5,3-.5) --++(-.5,0);
    \draw[thick] (-.5,4-.5) --++(-.5,0);
    \draw[thick] (-.5,5-.5) --++(-.5,0);
    \draw[very thick, color=gray] (-.75, -.5) --++ (0, 5);
    \draw[thick, fill=black] (-.75, -.5) circle (1pt); 
    \draw[thick, fill=black] (-.75, 1-.5) circle (1pt); 
    \draw[thick, fill=black] (-.75, 2-.5) circle (1pt); 
    \draw[thick, fill=black] (-.75, 3-.5) circle (1pt); 
    \draw[thick, fill=black] (-.75, 4-.5) circle (1pt); 
    \draw[thick, fill=black] (-.75, 5-.5) circle (1pt); 
    \node[scale=1.5] at (-.75,5) {$L^{(2T)}$};
}=
\fineq[0.8ex][.7][1]{
    \roundgate[0][0][1][topright][bertiniredstronger]
    \roundgate[1][1][1][topright][bertiniredstronger]
    \roundgate[0][2][1][topright][bertiniredstronger]
    \roundgate[1][3][1][topright][bertiniredstronger]
    \roundgate[0][4][1][topright][bertiniredstronger]
    \draw[thick] (-.5,-.5) --++(-.5,0);
    \draw[thick] (-.5,1-.5) --++(-.5,0);
    \draw[thick] (-.5,2-.5) --++(-.5,0);
    \draw[thick] (-.5,3-.5) --++(-.5,0);
    \draw[thick] (-.5,4-.5) --++(-.5,0);
    \draw[thick] (-.5,5-.5) --++(-.5,0);
},
\qquad \text{ where } L^{(2T)} = V \left(I^{\otimes (T-1)} \otimes l_{P(\alpha)}\otimes l_{\alpha}\right) V^\dag
\end{aligned}
\ee
Now, if one looks at the elements of $R$ that emerge in the expressions in Eq.~\eqref{eq:app_local_graphical_conditions_CDU2}, one can easily see that their generators commute with $O$ and therefore with $V$. For this reason, we find that the element of $R$ that was not dealt with at the first column anticommutes with $L^{(2t)}$. Therefore, all elements of $R$ are corrected by the two column stabilisers such that the larger local conditions hold
\be
\label{eq:app_local_graphical_conditions_CDU2}
\begin{aligned}
\fineq[-.8ex][.6][1]{
    \roundgate[0][0][1][topright][\col]
    \roundgate[1][1][1][topright][\col]
    \roundgate[0][2][1][topright][\col]
    \roundgate[1][3][1][topright][\col]
    \roundgate[2][2][1][topright][\col]
    \cstate[.5][3.5]
    \cstate[-.5][2.5]
    \cstate[-.5][1.5]
    \cstate[-.5][.5]
    \cstate[-.5][-.5]
} = 
\fineq[-.8ex][.6][1]{
    \roundgate[0][0][1][topright][\col]
    \roundgate[1][1][1][topright][\col]
    \roundgate[1][3][1][topright][\col]
    \roundgate[2][2][1][topright][\col]
    \cstate[.5][3.5]
    \cstate[.5][2.5]
    \cstate[.5][1.5]
    \cstate[-.5][.5]
    \cstate[-.5][-.5]
}, \quad \text{for } v_0=\frac{1}{3}, \qquad
\fineq[-.8ex][.6][1]{
    \roundgate[0][0][1][topright][\col]
    \roundgate[1][1][1][topright][\col]
    \roundgate[0][2][1][topright][\col]
    \roundgate[1][3][1][topright][\col]
    \roundgate[0][4][1][topright][\col]
    \roundgate[1][5][1][topright][\col]
    \roundgate[2][2][1][topright][\col]
    \roundgate[2][4][1][topright][\col]
    \cstate[.5][5.5]
    \cstate[-.5][4.5]
    \cstate[-.5][3.5]
    \cstate[-.5][2.5]
    \cstate[-.5][1.5]
    \cstate[-.5][.5]
    \cstate[-.5][-.5]
} = 
\fineq[-.8ex][.6][1]{
    \roundgate[0][0][1][topright][\col]
    \roundgate[1][1][1][topright][\col]
    \roundgate[0][2][1][topright][\col]
    \roundgate[1][3][1][topright][\col]
    \roundgate[1][5][1][topright][\col]
    \roundgate[2][2][1][topright][\col]
    \roundgate[2][4][1][topright][\col]
    \cstate[.5][5.5]
    \cstate[.5][4.5]
    \cstate[.5][3.5]
    \cstate[-.5][2.5]
    \cstate[-.5][1.5]
    \cstate[-.5][.5]
    \cstate[-.5][-.5]
},\quad \text{for } v_0=\frac{1}{5} \text{ and so on ...}
\end{aligned}
\ee
One may now use these graphical equations to prove the CDU3 conditions with the only non-triviality being that one must use certain subequations of the above found by contracting with the identity on 

To conclude, we now combine the deformations described here with those summarised by Property \ref{prop:cdu2_general}. We consider a free fermionic Clifford gate subject to the same requirements as above but now deformed according to Property \ref{prop:cdu2_general} with the extra restriction that $u_3=u_4=I$. That is we take our gate to be
\be
    W_{x,x+1/2} = (u_1\otimes u_2) \,\, U_{x,x+1/2} \exp(i\theta \gamma_x^{(1)}\gamma_{1,x}^{(v_0)}\gamma_{1,x+1/2}^{(v_0)}\gamma_{x+1/2}^{(1)})
\ee
Now, we know from Property \ref{prop:cdu2_general} that these single-site unitaries do not affect the stabilisers of the gate and, as such, the discussion above holds until we want to consider deforming the four Majorana term through the network. At this point they in general become problematic and we must add further restrictions. We will simply state the restrictions and then demonstrate why they work.

The first restriction is that they commute with the single problematic error that goes uncorrected at the one-column level Eq.~\eqref{eq:app_problematic_R}. This brings a restriction on to $u_1$ that 
\be
\label{eq:app_u1_cdu3_restrictions}
    [u_1, r_{\alpha_{2n-1}}] = 0, \quad n=1,\dots,\lfloor T/2 \rfloor.
\ee
The second is that we must be able to still deform the Majoranas through the network for which we impose upon $u_2$
\be
\label{eq:app_u2_cdu3_restrictions}
    [u_2, \ell_{\beta_{2n-1}}] = 0, \quad n=1,\dots,\lfloor T/2 \rfloor. 
\ee
It is interesting to note the symmetry of these equations: 
commuting with the problematic error and commuting with the deformed interaction are identical statements when we flip left to right. Another point of note is that there appears to be an increasing degree of restriction applied to the single site gates as $T$ increases. This however should be seen as counterbalanced by the implicit larger local Hilbert space dimension required for larger $T$. Furthermore, one can easily see elements of the algebra that contain $u_1$ and $u_2$ will often have elements that can survive any number commutations with stabilisers. For example, any quadratic form of $\ell$'s or $r$'s corresponding to fermions moving at $v=\pm1$ or more broadly $v\neq v_0$. The logic of these conditions is that, representing $u_1$ with a blue circle and $u_2$ with a red circle, we may make the following steps
\be
\begin{aligned}
\fineq[0.8ex][.7][1]{
    \roundgate[0][0][1][topright][bertiniredstronger]
    \roundgate[1][1][1][topright][bertiniredstronger]
    \roundgate[0][2][1][topright][bertiniredstronger]
    \roundgate[1][3][1][topright][bertiniredstronger]
    \roundgate[0][4][1][topright][bertiniredstronger]
    \draw[thick] (-.5,-.5) --++(-.5,0);
    \draw[thick] (-.5,1-.5) --++(-.5,0);
    \draw[thick] (-.5,2-.5) --++(-.5,0);
    \draw[thick] (-.5,3-.5) --++(-.5,0);
    \draw[thick] (-.5,4-.5) --++(-.5,0);
    \draw[thick] (-.5,5-.5) --++(-.5,0);
    \draw[thick] (.5,5-.5) --++(.5,0);
    \draw[thick] (.5,-.5) --++(.5,0);
    \cstate[-.5][.5][][bertiniblue]
    \cstate[.5][.5][][bertinired]
    \cstate[1-.5][1+.5][][bertiniblue]
    \cstate[1+.5][1+.5][][bertinired]
    \cstate[0-.5][2+.5][][bertiniblue]
    \cstate[0+.5][2+.5][][bertinired]
    \cstate[1-.5][3+.5][][bertiniblue]
    \cstate[1+.5][3+.5][][bertinired]
    \cstate[0-.5][4+.5][][bertiniblue]
    \cstate[0+.5][4+.5][][bertinired]
} \rightarrow
\fineq[0.8ex][.7][1]{
    \roundgate[0][0][1][topright][bertiniredstronger]
    \roundgate[1][1][1][topright][bertiniredstronger]
    \roundgate[0][2][1][topright][bertiniredstronger]
    \roundgate[1][3][1][topright][bertiniredstronger]
    \roundgate[0][4][1][topright][bertiniredstronger]
    \draw[thick] (-.5,-.5) --++(-.5,0);
    \draw[thick] (-.5,1-.5) --++(-.5,0);
    \draw[thick] (-.5,2-.5) --++(-.5,0);
    \draw[thick] (-.5,3-.5) --++(-.5,0);
    \draw[thick] (-.5,4-.5) --++(-.5,0);
    \draw[thick] (-.5,5-.5) --++(-.5,0);
    \draw[thick] (.5,5-.5) --++(.5,0);
    \draw[thick] (.5,-.5) --++(.5,0);
    \cstate[.5][.5][][bertinired]
    \cstate[1-.5][1+.5][][bertiniblue]
    \cstate[1+.5][1+.5][][bertinired]
    \cstate[0+.5][2+.5][][bertinired]
    \cstate[1-.5][3+.5][][bertiniblue]
    \cstate[1+.5][3+.5][][bertinired]
    \cstate[0+.5][4+.5][][bertinired]
}
=\fineq[-.6ex][.7][1]{
    \roundgate[0][0][1][topright][bertinired]
    \roundgate[1][1][1][topright][bertiniredstronger]
    \roundgate[0][2][1][topright][bertiniredstronger]
    \roundgate[1][3][1][topright][bertiniredstronger]
    \roundgate[0][4][1][topright][bertiniredstronger]
    \draw[thick] (-.5,-.5) --++(-.5,0);
    \draw[thick] (-.5,1-.5) --++(-.5,0);
    \draw[thick] (-.5,2-.5) --++(-.5,0);
    \draw[thick] (-.5,3-.5) --++(-.5,0);
    \draw[thick] (-.5,4-.5) --++(-.5,0);
    \draw[thick] (-.5,5-.5) --++(-.5,0);
    \draw[thick] (.5,5-.5) --++(.5,0);
    \draw[thick] (.5,-.5) --++(.5,0);
    \cstate[.5][.5][][bertinired]
    \cstate[1-.5][1+.5][][bertiniblue]
    \cstate[1+.5][1+.5][][bertinired]
    \cstate[0+.5][2+.5][][bertinired]
    \cstate[1-.5][3+.5][][bertiniblue]
    \cstate[1+.5][3+.5][][bertinired]
    \cstate[0+.5][4+.5][][bertinired]
    \draw[very thick, color=gray] (-.75, -.5) --++ (0, 5);
    \draw[thick, fill=black] (-.75, -.5) circle (1pt); 
    \draw[thick, fill=black] (-.75, 1-.5) circle (1pt); 
    \draw[thick, fill=black] (-.75, 2-.5) circle (1pt); 
    \draw[thick, fill=black] (-.75, 3-.5) circle (1pt); 
    \draw[thick, fill=black] (-.75, 4-.5) circle (1pt); 
    \draw[thick, fill=black] (-.75, 5-.5) circle (1pt); 
    \node[scale=1.5] at (-.75,5) {$V$};
}
\end{aligned}
\ee
where in the first step we used that the $u_1$ gates (blue circles) commute with the elements of $R$ unaccounted for by the first column,  i.e. Eq.~\eqref{eq:app_u1_cdu3_restrictions}, in order to remove them as they will trivially cancel when conjugating those elements of $R$. The second equality follows by the same logic of Eq.~\eqref{eq:app_majorana_deformation} but now noting that the $u_1$ gates (blue circles) trivially commute through the deformed Majoranas due to their definition in Property \ref{prop:cdu2_general} and that the $u_2$ gates (red circles) commute by the additional restrictions we have added in Eq.~\eqref{eq:app_u2_cdu3_restrictions}. At this point the proof follows as before.

\end{document}